\documentclass[twocolumn]{aastex631}
\usepackage{longtable}
\usepackage{catchfile}
\usepackage{tablefootnote}
\usepackage{footnote}
\usepackage{amsmath}
\usepackage[normalem]{ulem}
\shorttitle{Long-term monitoring of PSR~J1846$-$0258}
\shortauthors{R. Sathyaprakash et al.}
\begin{document}

\title{Long-term study of the 2020 magnetar-like outburst of the young pulsar PSR\,J1846--0258 in Kes 75}
%% you desire. In v6.2 you can include a footnote in the title.

%% Use \email to set provide email addresses. Each \email will appear on its
%% own line so you can put multiple email address in one \email call. A new
%% \correspondingauthor command is available in V6.2 to identify the
%% corresponding author of the manuscript. It is the author's responsibility
%% to make sure this name is also in the author list.
%%

\correspondingauthor{Rajath Sathyaprakash}
 
\author[0000-0002-5254-3969]{R. Sathyaprakash}
\email{rajath.sathyaprakash@iusspavia.it}
\affil{Institute of Space Sciences (ICE, CSIC), Campus UAB, Carrer de Can Magrans s/n, 08193, Barcelona, Spain}
\affil{Institut d'Estudis Espacials de Catalunya (IEEC), Esteve Terradas 1, 08860, Castelldefels, Barcelona, Spain} 
\affil{Scuola Universitaria Superiore IUSS Pavia, Palazzo del Broletto, piazza della Vittoria 15, I-27100 Pavia, Italy}

%\author[0000-0002-0430-6504]{E. Parent}
%\affiliation{Institute of Space Sciences (ICE, CSIC), Campus UAB, Carrer de Can Magrans s/n, 08193, Barcelona, Spain}
%\affiliation{Institut d'Estudis Espacials de Catalunya (IEEC), Esteve Terradas 1, 08860, Castelldefels, Barcelona, Spain} 

\author[0000-0003-2177-6388]{N. Rea}
\affiliation{Institute of Space Sciences (ICE, CSIC), Campus UAB, Carrer de Can Magrans s/n, 08193, Barcelona, Spain}
\affil{Institut d'Estudis Espacials de Catalunya (IEEC), Esteve Terradas 1, 08860, Castelldefels, Barcelona, Spain} 

\author[0000-0001-7611-1581]{F. Coti Zelati}
\affiliation{Institute of Space Sciences (ICE, CSIC), Campus UAB, Carrer de Can Magrans s/n, 08193, Barcelona, Spain}
\affil{Institut d'Estudis Espacials de Catalunya (IEEC), Esteve Terradas 1, 08860, Castelldefels, Barcelona, Spain} 

\author[0000-0001-8785-5922]{A. Borghese}
\affiliation{Instituto de Astrofísica de Canarias (IAC), Vía Láctea s/n, La Laguna 38205, S/C de Tenerife, Spain}
\affiliation{European Space Astronomy Centre (ESA/ESAC), Villanueva de la Ca\~nada (Madrid), Spain} 

\author[0000-0001-7397-8091]{M. Pilia} 
\affiliation{INAF Osservatorio Astronomico di Cagliari, Via della Scienza 5, I-09047 Selargius, Italy}

\author[0000-0002-1530-0474]{M. Trudu} 
\affiliation{Università degli Studi di Cagliari, Dipartimento di Fisica, SP Monserrato-Sestu km 0.7, I-09042 Monserrato (CA), Italy}
\affiliation{INAF Osservatorio Astronomico di Cagliari, Via della Scienza 5, I-09047 Selargius, Italy}

\author[0000-0002-8265-4344]{M. Burgay} 
\affiliation{INAF Osservatorio Astronomico di Cagliari, Via della Scienza 5, I-09047 Selargius, Italy}

\author[0000-0003-3977-8760]{R. Turolla} 
\affiliation{Dipartimento di Fisica e Astronomia “Galileo Galilei”, Università di Padova, via F. Marzolo 8, I-35131 Padova, Italy}
\affiliation{Mullard Space Science Laboratory, University College London, Holmbury St Mary, Dorking, Surrey RH5 6NT, UK}
	
\author[0000-0001-5326-880X]{S. Zane} 
\affiliation{Mullard Space Science Laboratory, University College London, Holmbury St Mary, Dorking, Surrey RH5 6NT, UK}

\author[0000-0003-4849-5092]{P. Esposito} 
\affiliation{Scuola Universitaria Superiore IUSS Pavia, Palazzo del Broletto, piazza della Vittoria 15, I-27100 Pavia, Italy}
\affiliation{INAF—Istituto di Astrofisica Spaziale e Fisica Cosmica di Milano, via A. Corti 12, I-20133 Milano, Italy}

\author[0000-0003-3259-7801]{S.~Mereghetti} 
\affiliation{INAF—Istituto di Astrofisica Spaziale e Fisica Cosmica di Milano, via A. Corti 12, I-20133 Milano, Italy}

\author[0000-0001-6278-1576]{S.~Campana} 
\affiliation{INAF--Osservatorio Astronomico di Brera, Via Bianchi 46, Merate (LC), I-23807, Italy}

\author[0000-0001-9494-0981]{D.~G\"otz} 
\affiliation{AIM-CEA/DRF/Irfu/Département d’Astrophysique, CNRS, Université Paris-Saclay, Université de Paris Cité, Orme des Merisiers, F-91191 Gif-sur-Yvette, France}

\author[0000-0002-5663-1712]{A.Y. Ibrahim}
\affiliation{Institute of Space Sciences (ICE, CSIC), Campus UAB, Carrer de Can Magrans s/n, 08193, Barcelona, Spain}
\affiliation{Institut d'Estudis Espacials de Catalunya (IEEC), Esteve Terradas 1, 08860, Castelldefels, Barcelona, Spain} 

\author[0000-0001-5480-6438]{G. L. Israel} 
\affiliation{INAF—Osservatorio Astronomico di Roma, via Frascati 33, I-00078 Monteporzio Catone, Italy}

\author[0000-0001-5902-3731]{A. Possenti}
\affiliation{INAF Osservatorio Astronomico di Cagliari, Via della Scienza 5, I-09047 Selargius, Italy}

\author[0000-0002-6038-1090]{A. Tiengo}
\affiliation{Scuola Universitaria Superiore IUSS Pavia, Palazzo del Broletto, piazza della Vittoria 15, I-27100 Pavia, Italy}
\affiliation{INAF—Istituto di Astrofisica Spaziale e Fisica Cosmica di Milano, via A. Corti 12, I-20133 Milano, Italy}

\def\xmm {\emph{XMM--Newton}}
\def\cxo {\emph{Chandra}}
\def\nustar {\emph{NuSTAR}}
\def\rst {\emph{ROSAT}}
\def\swift {\emph{Swift}}
\def\nicer {\emph{NICER}}
\def\pks {Parkes}

\def\flux {\mbox{erg\,cm$^{-2}$\,s$^{-1}$}}
\def\lum {\mbox{erg\,s$^{-1}$}}
\def\nh {N_{\rm H}}
\def\kms  {\rm \ km \, s^{-1}}
\def\cms  {\rm \ cm \, s^{-1}}
\def\gs   {\rm \ g  \, s^{-1}}
\def\cmtre {\rm \ cm^{-3}}
\def\cmdue {\rm \ cm$^{-2}$}
\def\ss {\mbox{s\,s$^{-1}$}}
\def\chisq {$\chi ^{2}$}
\def\rchisq {$\chi_{r} ^{2}$}

\def\arc{\mbox{$^{\prime\prime}$}}
\def\arcmin{\mbox{$^{\prime}$}}
\def\deg{\mbox{$^{\circ}$}}

\def\rsun {~R_{\odot}}
\def\msun {~M_{\odot}}
\def\mdotav {\langle \dot {M}\rangle }

\def\uu {4U\,0142$+$614}
\def\ee {1E\,1048.1$-$5937}
\def\kes {1E\,1841$-$045}
\def\aa {1E\,1547$-$5408}
\def\axj {AX\,J1844$-$0258}
\def\rxs {1RXS\,J1708$-$4009}
\def\xte{XTE\,J1810$-$197}
\def\smc{CXOU\,J0100$-$7211\,}
\def\wes{CXOU\,J1647$-$4552}
\def\ea {1E\,2259$+$586}
\def\ctb{CXOU\,J171405.7$-$381031}
\def\sgra{SGR\,1806$-$20}
\def\sgrb{SGR\,1900$+$14}
\def\sgrd{SGR\,1627$-$41}
\def\sgre{SGR\,0501$+$4516}
\def\sgrf{SGR\,1935+2154}
\def\lowba{SGR\,0418$+$5729}
\def\sgrg{SGR\,1833$-$0832}
\def\lowbb{Swift\,J1822.3$-$1606}
\def\galmag{PSR~J1745$-$2900}
\def\sgras{Sgr\,A$^{\star}$}
\def\sgrh{SGR\,1801$-$21}
\def\sgri{SGR\,2013$+$34}
\def\psr{PSR\,1622$-$4950}
\def\hbpsr{PSR~J1846$-$0258}
\def\radiohb{PSR~J1119$-$6127}
\def\coronamag{Swift~J1818.0$-$1607}

\def\srcfirst {\mbox{Swift\,J1830.7$-$0645}}
\def\src {\mbox{Swift\,J1830}}

\begin{abstract}
Magnetar-like activity has been observed in a large variety of neutron stars. \hbpsr\ is a young 327\,ms radio-quiet pulsar with a large rotational power ($\sim 8 \times10^{36}$\,erg\,s$^{-1}$), and resides at the center of the supernova remnant Kes\,75. It is one of the rare examples of a high magnetic field pulsar showing characteristics both of magnetars and radio pulsars, and can thus provide important clues on the differences in the emission mechanisms between these two classes. In 2006, \hbpsr\ was detected to undergo an outburst for the first time, accompanied by a large flux increase, millisecond X-ray bursts, significant spectral changes and a large timing glitch. In the period between May-June 2020, after fourteen years of quiescent stable emission, the source underwent a second magnetar-like outburst, which was followed up with several observations by {\it{NICER}}, {\it{XMM-Newton}}, {\it{NuSTAR}} and {\it{Swift}}. In this work, we report on the long-term timing and X-ray spectral properties of the source following the 2020 outburst, and place upper limits on any source activity at radio wavelengths. We demonstrate that the pulsed flux increased by a factor $> 6$ during the outburst, followed by non-trivial variability in the spin-down rate. Our timing analysis shows that the spin frequency and its derivative are clearly affected by magnetospheric activity due to the outburst. We find hints for an oscillation in the frequency derivative with a timescale of 50--60\,days, recovering later on to stable quiescence. 
\end{abstract}

\section{Introduction} \label{sec:intro}

Isolated neutron stars (NSs) are the compact remnants of core-collapse supernova explosions, characterised by fast rotation velocities and large magnetic fields. Observational surveys limited to the Milky Way alone have discovered more than three thousand isolated NSs in the past few decades. A handful of around 30 such objects, known as magnetars, were historically observed as flaring X-ray and gamma-ray sources and are believed to be powered by extreme magnetic fields (i.e. $\sim 10^{14}-10^{15}$\,G) (for a detailed review see e.g. \citealt{esposito21} and \citealt{kaspi17}) with relatively slow spin periods of 1-10\,s and spin-down ages ranging from $10^{2} - 10^{4}$\,yr. Among the key attributes of magnetars is that their X-ray luminosity appears to far exceed the available spin-down power, especially during periods of short duration (i.e. milli-second) bursts and long duration outbursts. Their spectra during quiescence are typically modelled by blackbody emission with temperatures $\sim 0.3-1.0$ keV, with a hard X-ray tail. 

%------- FIGURE ---------
\begin{figure*}
\centering
\includegraphics[scale=0.65]{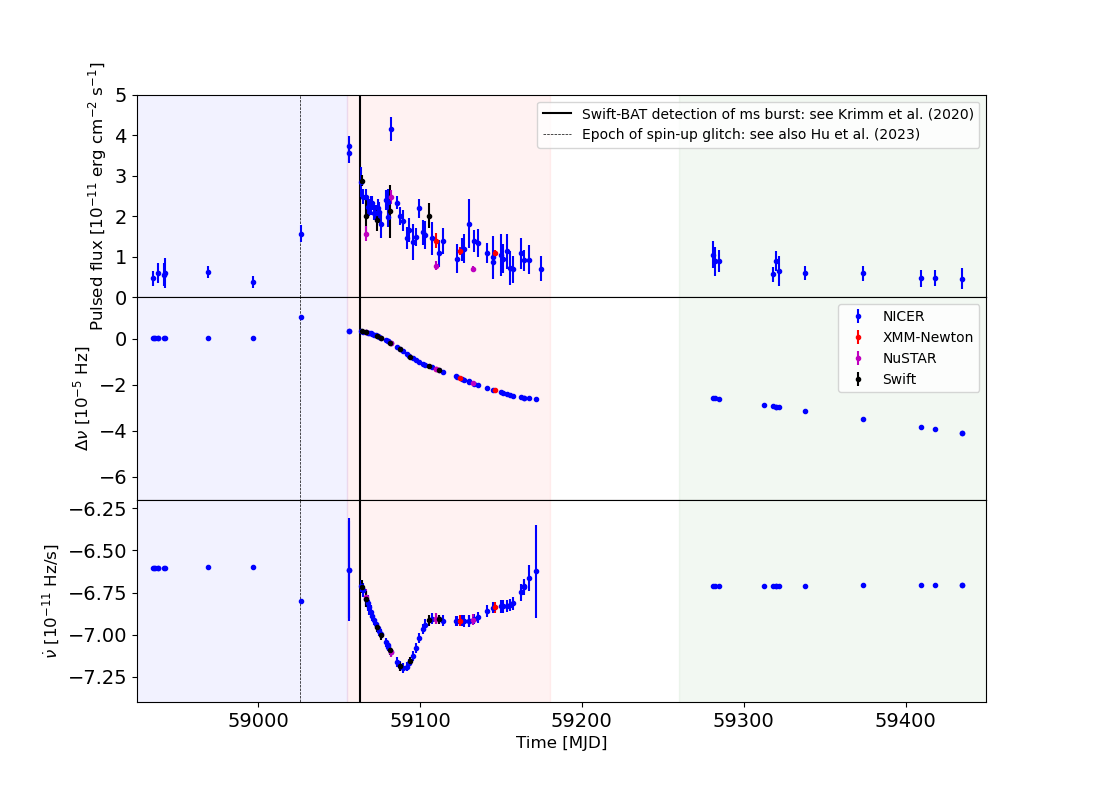}
\caption{Top panel: evolution of the pulsed flux (in the 1 - 12 keV energy range) covering the periods before (left blue shaded region), during (red shaded region) and after (right green shaded region) the August 2020 outburst of \hbpsr.\ The solid black line denotes the epoch when the {\it{Swift}} Burst Alert Telescope (BAT) and Fermi Gamma-Ray Burst Monitor (GBM) both confirmed the detection of a short ($<100$ milli-second) burst from the source, while the dashed black line indicates a slightly earlier period during which a spin-up glitch/irregularity was found to occur with respect to the pre-outburst spin evolution. The pulsed flux was computed using phase resolved spectroscopy after deriving the spin parameters $\nu$ and $\dot{\nu}$ via the tempo-derived phase-coherent timing approach (see Section~\ref{sec:coherent-timing}) and Gaussian process regression (GPR) (see Section~\ref{sec:GPR}) Middle panel: Evolution in the spin frequency relative to the pre-outburst timing solution (i.e. before MJD 59026) $\Delta \nu$ derived from computing the derivative of the best-fit Gaussian process model to the timing residuals. Bottom panel: Evolution in the frequency derivative $\dot{\nu}$. For all three panels, the errors are quoted at the 1$\sigma$ confidence level}.
\label{Fig:main_plot}
\end{figure*}
%------------------------

The difference between the emission of rotation-powered pulsars and magnetars has been studied for many years. Two decades ago, the former were observed mainly as stable radio emitters, while the latter were observed as radio-quiet bright X-ray sources. This was attributed to a distinct emission reservoir and physics between the two classes. The discovery of radio-emitting magnetars \citep{camilo06}, or magnetar-like emission in allegedly canonical radio pulsars (e.g. \citealt{gavriil08,archibald16}) and low magnetic field magnetars \citep[e.g.][]{rea10} has completely changed our understanding of the magnetic-powered emission across the isolated NS zoo. Indeed, more recent studies show that the population of isolated magnetars appears to not only be restricted to Galactic sources, and those that have been observed to emit milli-second duration radio flares have been posited to be excellent candidates for fast radio bursts (e.g. \citealt{wang2022}), a few of which have been deduced to be extra-galactic in origin due to their large dispersion measures. 

Located within the core of the supernova remnant (SNR) Kes\,75, \hbpsr\ (first discovered by \citealt{gotthelf20}), is one of the youngest pulsars in our Galaxy with a characteristic age $\sim 700$ years, spin period of 327\,ms and a distance of 5.8 kpc\footnote{derived from analysing HI and $^{13}$CO maps of the Kes-75 remnant (\citealt{leahy08})} (\citealt{verbiest2012}). It has a dipolar magnetic field of $B \sim 4.5 \times 10^{13}$\,G and has historically been radio-quiet but a very powerful X-ray emitter. Similar to other rotation-powered pulsars, it has a large rotational power ($\dot{E} \sim 8 \times 10^{36}$ erg\,s$^{-1}$) and lies at the center of a bright pulsar wind nebula (PWN).

For many years prior to 2006, it behaved as a very stable X-ray pulsar. However, in 2006, the source displayed a distinctly magnetar-like behaviour, undergoing a dramatic increase in pulsed flux that coincided with the detection of X-ray bursts lasting a few milliseconds \citep{gavriil08}, showed significant spectral changes \citep{kumar08} and a large timing glitch \citep{kuiper09a}. Therefore, this object has been recognized as an interesting hybrid between magnetars and rotation-powered pulsars that helps to understand the links between these two different classes of NSs. 

In 2009, a {\it{Chandra}} observation caught the source in quiescence \citep{livingstone11a}, with a braking index ($n = 2.16 \pm 0.13$) that differed significantly from that measured during the pre-outburst epoch ($n=2.65 \pm 0.01$) in 2006 \citep{livingstone06}. Indeed, further investigations showed that the braking index had changed permanently following the magnetar-like outburst and its radiatively loud glitch \citep{archibald15a}. Thus far, no physical model can consistently explain the long-term decrease in the braking index.

Other works have focused on studying the emission of the pulsar and its wind nebula using {\it{XMM-Newton}}, {\it{NuSTAR}} and {\it{Chandra}} observations (e.g., \citealt{gotthelf21}). By performing phase-resolved spectroscopy, \citealt{gotthelf21} found that the joint {\it{XMM-Newton}} and {\it{NuSTAR}} spectrum of the pulsar is characterised by a power-law model in the 2-50 keV energy band. However, when analyzing a broader energy range (2 keV--100 MeV) using data from {\it{RXTE}}, {\it{INTEGRAL}} and {\it{Fermi}}, \cite{kuiper18} found evidence for the presence of spectral curvature in the pulsed emission of \hbpsr. This phenomenology could potentially be explained in the context of the Outer Gap, Slot Gap or Polar Cap models, in which non-thermal gamma-ray photons are emitted due to the synchrotron and inverse Compton processes by accelerating electrons. \cite{kuiper18} also reported weak pulsed emission in the soft gamma-ray range (30 - 100 MeV) analysing {\it{Fermi-LAT}} data collected during a $\sim 8$ year period, after the source had returned to a quiescent state following its 2006 outburst. In more recent work, \cite{straal23} used {\it{Fermi}}-LAT data to observe the spectral energy distribution of the source beyond 100 MeV, and thereby placed constraints on the properties of the PWN (i.e. its magnetisation and particle energy spectrum), in addition to the birth properties of the NS and supernova ejecta.  

In the period between May-June 2020, the source underwent a second outburst (followed by a short 100 ms duration burst in August 2020; see \citealt{krimm20} and \citealt{uzuner2023}) after a long quiescent period that had lasted for more than 14 years \citep{krimm20,laha20}. This event triggered the onset of several observations with the Neutron Star Interior Composition Explorer ({\it{NICER}}) and the Neil Gehrels {\it{Swift}} Observatory ({\it{Swift}}). These observations revealed the occurrence of a spin-up glitch (\citealt{kuiper20}) accompanied by a significant increase in pulsed flux, reaching over five times the quiescent level, with the pulsed spectrum characterised by the combination of a blackbody and a power-law component between 1 - 70 keV \citep{hu23}. Spectral analysis indicated that the flux increase was primarily caused by the emergence of the thermal component, which gradually faded away, while the power-law component remained constant throughout the outburst \citep{hu23}. Observations with the Deep Space Network (DSN) 34 meter radio instrument early in the outburst resulted in a non-detection of radio pulsations from the source, consistent with previous observations (see \citealt{blum21}, \citealt{majid20} and \citealt{mickaliger2020}).

In the present work, we report on the long-term timing and spectral properties of the source following the recent outburst using four {\it{NuSTAR}} observations, three {\it{XMM-Newton}} observations and eight {\it{Swift}} observations, all triggered in the aftermath of the milli-second burst (see \citealt{krimm20}). We also include several {\it{NICER}} observations obtained over three years from August 2018 until August 2021 (see Table~\ref{Tab:Table5} and \citealt{hu23}).

\section{Data reduction} 
\label{sec:Reduction}

\begin{figure}
\vspace{0.5cm}
\begin{center}
\includegraphics[scale=0.25]{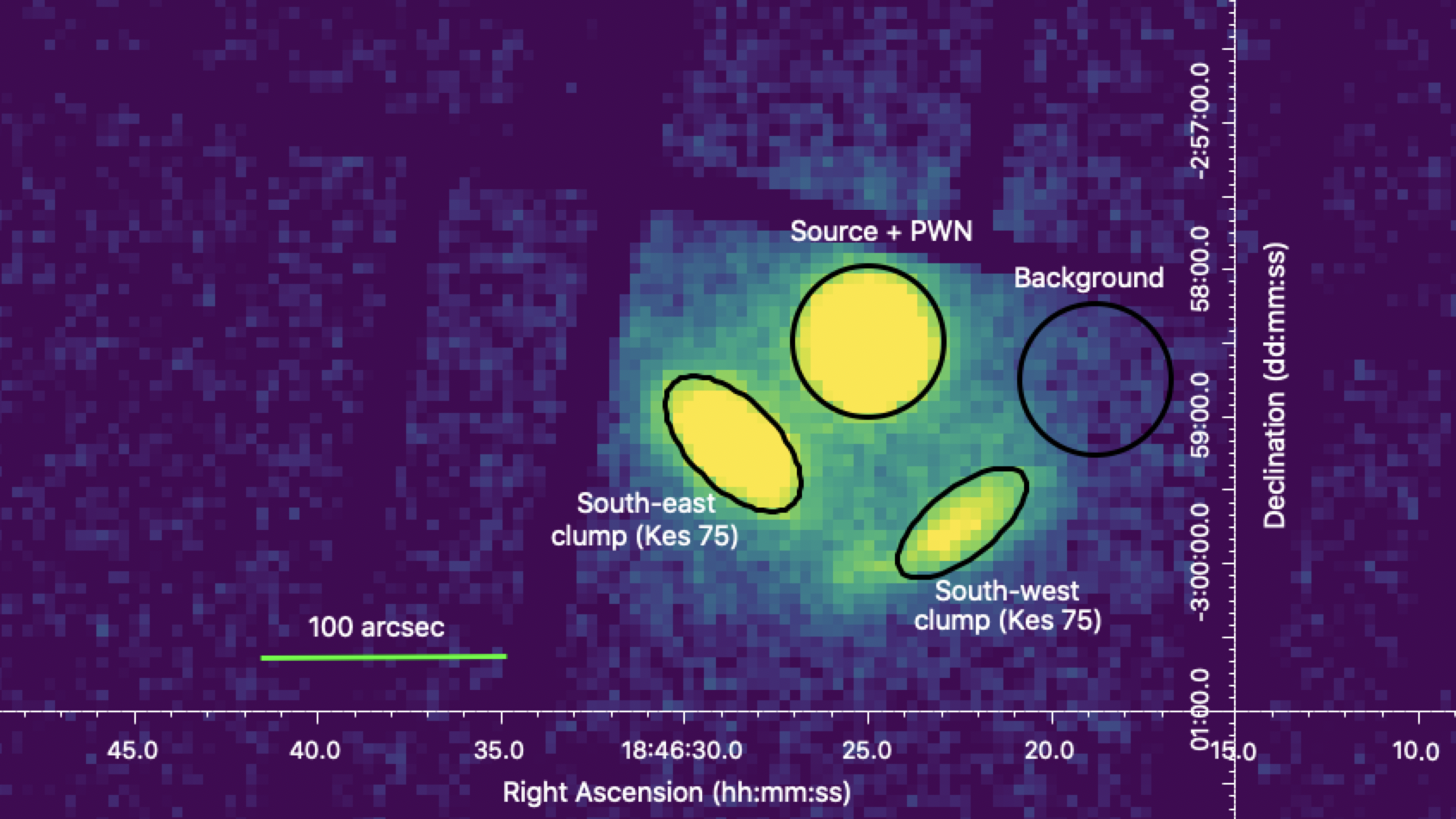}
\end{center}
\caption{The {\xmm} EPIC-pn image of the field containing the high-B pulsar \hbpsr,\, its PWN and the spatially resolved supernova remnant Kes-75. The circular black solid line indicates the extraction region used for the phase-resolved spectral analysis of the source.}
\label{Fig:PNimage}
\end{figure}

The X-ray observations analysed in this work are listed in Table~\ref{Tab:Table5}. The EPIC-pn {\xmm} data were acquired in large window mode, while the EPIC-MOS data were collected in full-frame mode. Owing to their insufficient timing resolution (2.6 s), data from the MOS detectors were excluded for timing analysis. To process the {\xmm} data, we used the Science Analysis Software ({\tt{SAS v.18}}). We extracted a light-curve between 10--12\,keV for each observation (in the entire field-of-view) using {\tt{evselect}} to identify and filter out periods of intense background flares. We then generated spatially filtered EPIC-pn event files, extracting source events within a circular region of 30 arcseconds in the energy range of 1--12\,keV, centered on the most precise available {\cxo} source position (RA=281.6039167$^{\circ}$, DEC=-2.9750278$^{\circ}$; \citealt{helfand03}). The same position was adopted for all the other instruments. Background events were accumulated in a neighbouring circular region of 30 arcsecond radius, as illustrated in Figure~\ref{Fig:PNimage}. 

The {\nustar} data were processed with the {\tt{nupipeline}} script and source events were collected from a circular region of 50 arcsecond radius, within the recommended energy range of 3--79\,keV. Background photons were extracted from a neighbouring source-free circular region of the same size. The {\tt{nupipeline}} script excludes passages of the spacecraft through the South Atlantic Anomaly (SAA), with a recently updated algorithm that simultaneously monitors the count-rates from the high gain shield and a more efficient Cadmium-Zinc-Telluride (CTZ) detector, providing a cleaned event file that is suitable for timing analysis. 

The {\nicer} data were calibrated with the {\tt{nicerl2}} pipeline using the {\tt{heasoft}} package (version 6.31; \citealt{heasoft14}), which excludes time intervals within the SAA, in addition to overshoot and undershoot events caused by the charged particle background and solar optical photons interacting with the detector. Information from each Measurement Power Unit (MPU) was combined to produce a cleaned event file. 

Finally, the {\swift}-XRT data were acquired in windowed timing mode (WT; timing resolution of 1.8\,ms). We reprocessed the data adopting standard cleaning criteria with the task {\tt xrtpipeline}. The source photons were accumulated from a 10-pixel circular region (1 pixel=2$\farcs$36), while an annulus with inner and outer radii of 60 and 100 pixels, respectively, was used to extract the background events.

We barycentered event files from all instruments using the {\tt{barycorr}} tool available as part of the {\tt{heasoft}} package (v6.31), with the source position adopted from the {\cxo} observations (see above) and the latest DE-430 solar system ephemeris. For the {\xmm} data, we used the {\tt{barycen}} tool to perform the barycentric corrections. For the reasons discussed in Sec.~\ref{sec:pulse_profile}, we refer to all observations after 25th July 2020 (MJD 59055) as the post-outburst period, and observations before this date as the pre-outburst period. 

\section{Timing analysis}
\label{sec:Timing_analysis}

%------- FIGURE ---------
\begin{figure*}
\centering
\includegraphics[scale=0.55]{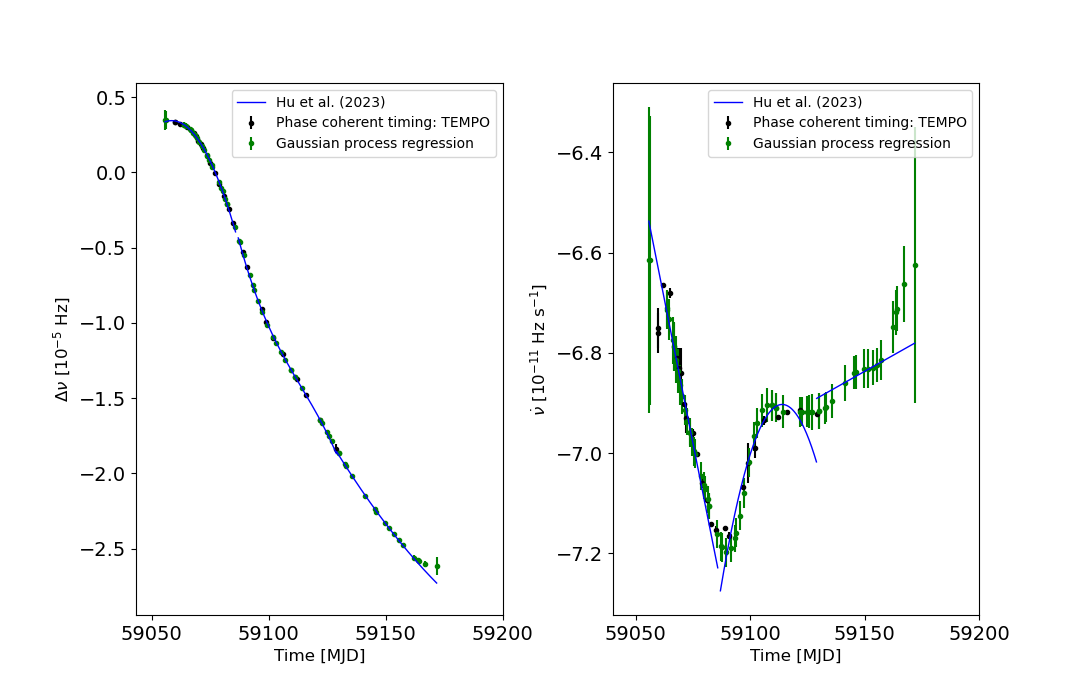}
\caption{Comparison between the values of the spin frequency $\Delta \nu$ (relative to the pre-outburst timing model; left) and the spin-down rate $\dot{\nu}$ (right) obtained in our study through phase-coherent timing (black points) and Gaussian process regression (GPR). We also show the values reported by \citealt{hu23} (solid blue line).}
\label{Fig:spin_par_comp}
\end{figure*}
%------------------------

During its 2006 outburst, \hbpsr\, experienced a spin-up glitch event \citep{kuiper09, livingstone11a} that coincided with a pulsed flux increase \citep{gavriil08} and the detection of four millisecond X-ray bursts. This was followed by a period of over-recovery, such that the glitch could not be described by a simple exponentially decaying model (\citealt{livingstone11a}). The magnitude of the over-recovery was larger than ever observed in other rotation-powered pulsars (RPPs) or magnetars, indicating that the model of vortex unpinning in the superfluid crust of the NS invoked to explain standard glitches in RPPs (see e.g. \citealt{haskell15}) likely does not hold for \hbpsr\, (see \citealt{livingstone11a} for more details). Here, we describe our timing analysis of \hbpsr\, throughout its 2020 outburst to monitor its spin evolution.

\subsection{Phase-coherent timing} 
\label{sec:coherent-timing}

To measure the spin parameters of \hbpsr\, we attempted to derive a coherent timing solution using the standard phase-connection approach carried out with pulsar timing software. We first assigned a rotational phase to each photon based on a provisional ephemeris using the {\tt{photonphase}} tool of the {\tt{PINT}} software package \citep{luo21}. The events for each observation were folded to produce integrated pulse profiles, with the adopted energy range discussed in the previous section. The pulse profile that had the highest $H$-test score was then fitted with a multi-Gaussian model to produce a standard profile template. We then used the {\tt{photon\_toa.py}} tool of the {\tt{NICERsoft}} package\footnote{\url{https://github.com/paulray/NICERsoft}} to extract as many reliable barycentered pulse time of arrivals (TOAs) as possible within each observation to monitor better the pulsar rotations during the phase-connection process. Initially, the only fitting parameters of the timing model were the spin frequency $\nu$, the first frequency derivative $\dot{\nu}$ and a phase offset, or ``jump'', for every gap between groups of TOAs from a same observation. 

We then proceeded with the phase connection with the {\tt{TEMPO}} pulsar timing software \citep{nds+15} by determining the rotation count between groups of TOAs (i.e. eliminating jump parameters) recursively from the shortest to longest gaps. The presence of strong timing noise in \hbpsr\, during its outburst required that we include the spin-frequency second derivative $\ddot{\nu}$ in the model early in the connection process in order to maintain coherence. As we reached the longest gaps, the rotation count became ambiguous: multiple statistically acceptable (reduced $\chi^{2} < 2$) and distinct solutions were found when connecting TOAs across these long gaps.

This was problematic in particular during the first two months of the outburst (2020 June and July) -- only three {\it{NICER}} observations of \hbpsr\, were carried out, and we find evidence of (unconstrained) timing anomalies as reported by \cite{hu23}. We attempted to eliminate phase ambiguities using the Dracula algorithm \citep{fr18}, but were unable to converge to a single global ephemeris.

To better monitor the spin-down behavior during the outburst, we instead opt to measure local phase-connected timing solutions with {\tt{TEMPO}}. TOAs from consecutive observations spanning only short time range of a few days were used for each such that the solutions were only sensitive to the $\nu$ and $\dot{\nu}$ parameters. Local solutions were produced from data taken $\sim$ 5 months prior to the onset of the outburst (i.e. June-August 2020) until one year after the outburst. Our best-fit $\nu$ and $\dot{\nu}$ measurements are shown as black points in Figure~\ref{Fig:spin_par_comp}. 

A timing analysis of \hbpsr\ during its 2020 outburst was also carried out by \cite{hu23} using {\it{NICER}}/XTI and {\it{Swift}}/XRT monitoring data -- their results are also shown in Figure~\ref{Fig:spin_par_comp} (solid blue line). We compare our measurements of the spin frequency and spin-down rate with the solutions from \cite{hu23}, which are generally in agreement with their values excluding a few observations (see Figure~\ref{Fig:spin_par_comp}). \cite{hu23} detected a strong glitch event of fractional size $\Delta{\nu}/\nu\sim3\times10^{-6}$ early during the outburst, between the epochs MJD 58996 and 59055. Similarly to \cite{hu23}, we detect a timing irregularity that is consistent with a glitch of estimated magnitude $\Delta{\nu}/\nu\sim2.5-4\times10^{-6}$ during the first month of the outburst (MJD 58996 to 59026). 

In summary, we find \hbpsr\, experienced rapid torque fluctuations during its outburst. Prior to 2020 June, the spin-down rate was stable at $\dot{\nu}\sim-6.6\times10^{-11}\,$Hz s$^{-1}$. A large $\sim 10^{-5}$\,Hz increase in $\nu$ and a 3\% decrease in $\dot{\nu}$ was measured in the 2020 June 26 (MJD 59026) {\it{NICER}} data, while the spin measurements the following month (i.e. July 2020) appeared to have evolved towards values more similar to the pre-outburst $\nu$ and $\dot{\nu}$. Such behavior in spin and spin-down rate is consistent with the expected spin evolution of a pulsar experiencing a spin-up glitch event (see e.g. \citealt{hobbs10}) between 2020 May 27 and 2020 June 26. The sparse sampling in this period yields poor constraints on the glitch epoch and thus several statistically valid timing models corresponding to different fractional frequency jumps $\Delta\nu/\nu$ in the range of $\sim2$-$5\times10^{-6}$ -- of similar size to the glitch event that accompanied the 2006 outburst of \hbpsr\ by \cite{kuiper09a}.

Between 2020 August 2 and August 21 (MJDs 59063 to 59082), unlike typical post-glitch recovery, the spin-down rate steadily decreased until it reached a minimum of $\sim -7.1 \times 10^{-11}\,$Hz s$^{-1}$. Over the following weeks $\dot{\nu}$ increased again but at an irregular rate. Our results, shown in Figure~\ref{Fig:spin_par_comp}, are consistent with the spin-down rates reported by \cite{hu23}. Due to visibility constraints, the monitoring of \hbpsr\, with {\it{NICER}} were interrupted at the end of 2020 November. When observations resumed in 2021 March, the pulsar's torque was stable again, but with a spin-down rate $\dot{\nu}\sim-6.7\times10^{-11}\,$Hz s$^{-1}$ slightly lower than its pre-outburst value. 

%------- TABLE ---------
%\begin{deluxetable*}{ccccccccc}
%\caption{Post-fit timing parameters of the coherent timing solutions derived in the four data segments. The second column lists the reference epochs for the measurements of the spin parameters $\nu$, $\dot{\nu}$. Values in parentheses correspond to the 1$\sigma$ uncertainties reported by {\tt{PINT}}. The JPL DE430 planetary ephemeris$^a$ and the Barycentric Dynamical Time (TDB) standard were used to derived the solutions below. 
%\label{Tab:timing}}
%\input{timing-table}
%\begin{itemize}
%    \footnotesize
%    \item[$^a$] \url{https://ssd.jpl.nasa.gov/planets/eph\_export.html}
%\end{itemize}
%\end{deluxetable*}
%------------------------

\subsection{Gaussian process regression}
\label{sec:GPR}

The timing residuals corresponding to the post-outburst period (between MJD 59055-59171), and extracted in the manner described in the previous section, display a complex quasi-periodic structure superposed on the secular spin-down of the source. In order to model these residuals, we implemented Gaussian process regression (GPR) via the {\tt{scikit-learn}} library \citep{pedregosa11}. GPR is a non-parametric regression method that is useful in cases where it is not possible to make clear assumptions about a physical model that the data might satisfy. Given a set of data points ($x$,$y$), a GPR model can fit (and interpolate between) these data points with functions sampled from a multi-variate Gaussian distribution. These functions are characterised by a mean vector and a covariance matrix $k(x,x^{\prime})$, whose diagonal elements describe the variance in the dependent variable $y$ being modelled, and the off-diagonal elements describe the similarity of the function (or its shape) between any two points ($y(x),y(x^{\prime})$). There are several functional forms available for $k(x,x^{\prime})$ and the choice of which to use depends on the data being modeled. A commonly used kernel is the radial basis function kernel (RBF):
\begin{equation}
	k(x,x^{\prime}) = \sigma_{f}^{2} \exp\bigg[-\frac{(x- x^{\prime})^{2}}{2l^{2}}\bigg]~.
\end{equation}
It is characterised by two hyper-parameters\footnote{defined as hyper-parameters, since they're only used in training the regression model, rather than describing any physically relevant properties of the timing residuals}, $\sigma_{f}$ and $l$, where  $\sigma_{f}$ describes the vertical extent of the function and the length scale $l$ describes how quickly the correlation between two points reduces as their distance increase (larger $l$ implies a smoother function). The hyper-parameters of the model were fine-tuned within the {\tt{scikit-learn}} module, which minimises the log-likelihood on the available data. The best-fit GPR model can be used to generate predictions $\boldsymbol{f}$ and their associated uncertainty $\text{var} (\boldsymbol{f})$ as follows:
\begin{equation}
	\boldsymbol{f} = \boldsymbol{K_{*}^{T}} (\boldsymbol{K} + \sigma_{n}^2 \boldsymbol{I})^{-1}\boldsymbol{y}~,
\end{equation}

\begin{equation}
	\text{var}(\boldsymbol{f}) = \text{diag} \bigg[ \boldsymbol{K_{**}} - \boldsymbol{K_{*}^{T}}(\boldsymbol{K} + \sigma_{n}^2 \boldsymbol{I})^{-1}\boldsymbol{K_{*}} \bigg ]~,
\end{equation}
where $\boldsymbol{I}$ is the identity matrix, $\boldsymbol{y}$ are the observed data points (in our case the timing residuals), the elements of covariance matrices are $\boldsymbol{K} = k(x,x)$, $\boldsymbol{K_{*}} = k(x,x^{\prime})$, $\boldsymbol{K_{**}} = k(x^{\prime},x^{\prime})$ and the apex ``T'' indicates the transpose matrix. For more details on the GPR method and its applications in pulsar timing we refer the readers to \cite{rasmussen06a,brook16,rajwade22}. In our case, the timing residuals were best modelled with two RBF kernels in addition to a white-noise kernel that accounts for uncertainties $\sigma_{n}$ in the timing residuals. The optimised hyper-parameters were: $\sigma_{f_{1}}=0.254 \pm 0.01$ cycles, $l_{1}=11.5 \pm 0.3$ days, $\sigma_{f_{2}}=26.5^{+1}_{-2}$ cycles, $l_{2}=37.6^{+0.4}_{-1}$ days and $\sigma_{n}=(1.7 \pm 0.4) \times 10^{-2}$ cycles.

We remark that, although the optimised GPR model does not provide any physical insights, it provides a means to evaluate derivatives of the timing residuals, which can be used to compute the observable parameters $\nu$ and $\dot{\nu}$. The derivatives and their uncertainties were evaluated analytically using the formalism outlined by \cite{brook16}. The derived evolution in $\nu$ and $\dot{\nu}$ are indicated in the middle and bottom panels of Figure~\ref{Fig:main_plot} and in Figure~\ref{Fig:spin_par_comp}, suggesting that these values are generally consistent with those derived using the phase coherent timing approach in the previous section and by \citet{hu23}. 

Our analysis suggests that the onset of the X-ray outburst (in the period between May-June 2020) triggered a dramatic change in the spin parameters, especially with the magnitude of $\dot{\nu}$ showing a relative increase (i.e. $\frac{\Delta \dot{\nu}}{\dot{\nu}}$) of more than 7\%, before reaching values close to the pre-outburst level a few months later. Possible physical explanation for such a complex variation in the spin parameters will be discussed further in Section~\ref{sec:discussion}. 

%Moreover, the braking index ($n = \nu \ddot{\nu}/\dot{\nu}^2$) of the source and its evolution has been studied in detail by \cite{archibald15a}, who report that its long-term value $n = 2.19 \pm 0.03$ deviates significantly from $n = 3$ predicted for a rotating magnetic dipole in vacuum. We note that in our work the pre-outburst spin parameters suggest a braking index that is entirely consistent with the findings of \cite{archibald15a}, although this measurement is based on a much smaller baseline. Indeed, they discuss several reasons that could explain $n$ being smaller than the canonical value, including a particle wind that carries away angular momentum from the system, variation in the moment of inertia or the angle between the spin and the magnetic axis. However, none of these possibilities seems to explain the observed variability in the braking index. This was also discussed by \cite{archibald15a}, who note that their measurement is not consistent with the value reported by \cite{livingstone06} before the 2006 outburst ($n = 2.65 \pm 0.01$). \hbpsr\ is one of the few sources for which such a large change in $n$ has been observed, but there doesn't seem to be any obvious explanations for this occurrence.  

%------- FIGURE ---------
\begin{figure*}
\centering
\includegraphics[scale=0.5]{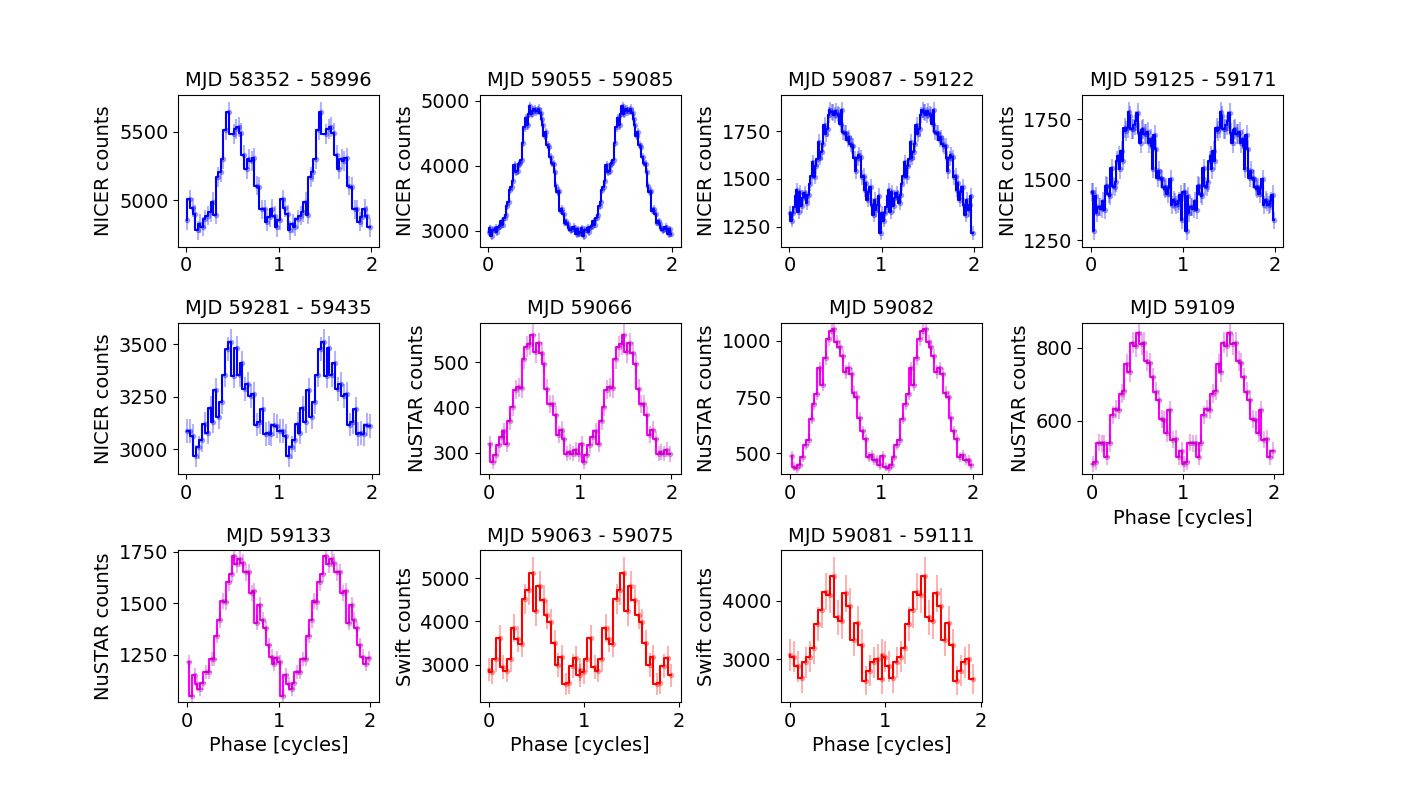}
\caption{Stacked {\it{NICER}} (blue), {\it{NuSTAR}} (magenta) and {\it{Swift}} (red) pulse profiles for the periods before (MJD 58352 - 58996), during (MJD 59055 - 59085, MJD 59087 - 59122 and MJD 59125 - 59171) and after (MJD 59281 - 59435) the 2020 outburst of PSR J1846-0258. The pulse profiles were combined over the observational epochs indicated at the top of each profile in the energy ranges of 1--12 keV (for {\it{NICER}}), 3--79 keV (for {\it{NuSTAR}}) and 1--10 keV for {\it{Swift}}.}
\label{Fig:nicer_pprof}
\end{figure*}
%------------------------

%------- FIGURE ---------
\begin{figure*}
\centering
\includegraphics[scale=0.5]{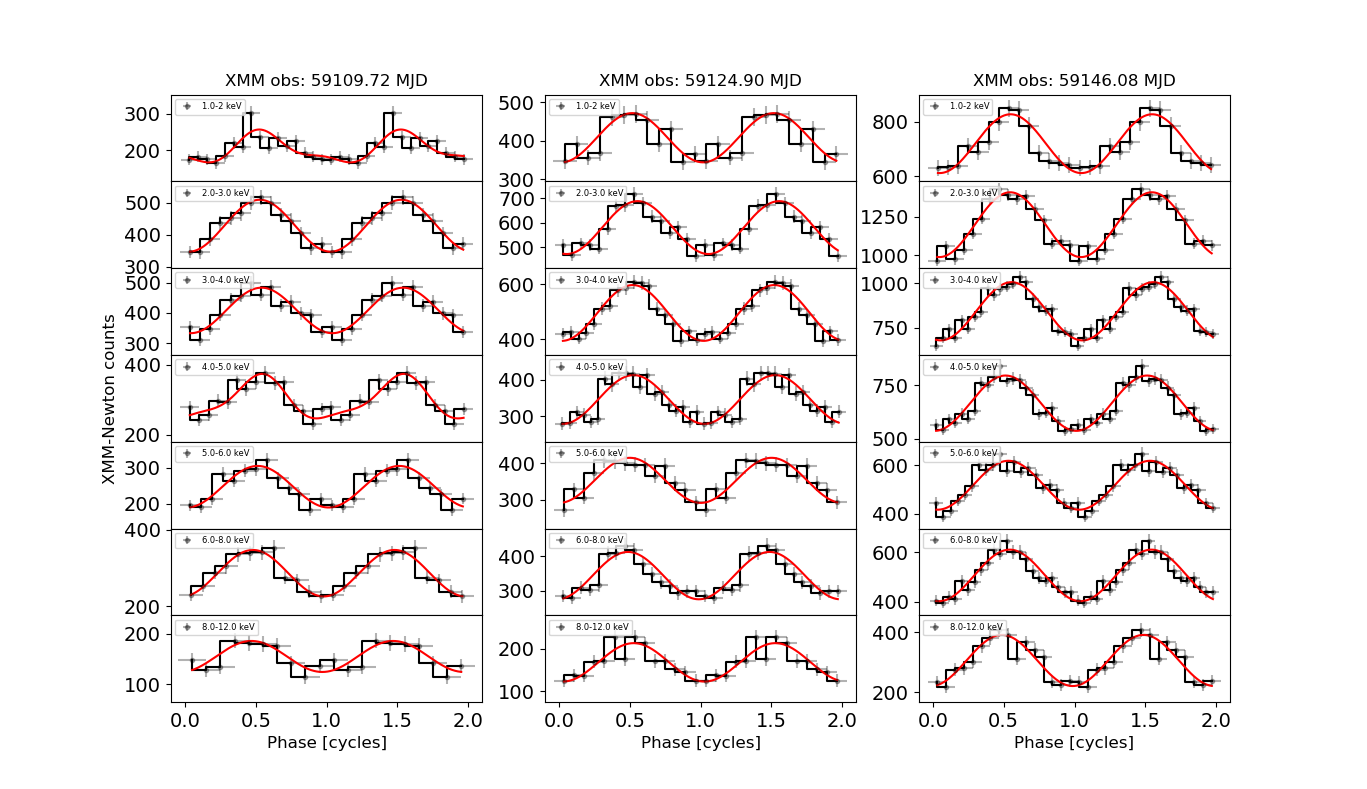}
\caption{Energy-resolved pulse profiles from the three post-outburst {\xmm} observations, with the red solid line displaying the best-fit sinusoidal model (with at-most two harmonic components). The number of phase bins for each profile was chosen based on the total number of counts available. Observational dates are indicated in the top panel of each plot.}
\label{Fig:xmm_pulse_prof}
\end{figure*}
%------------------------

%--------FIGURE ----------------
\begin{figure}
\centering
\includegraphics[scale=0.4]{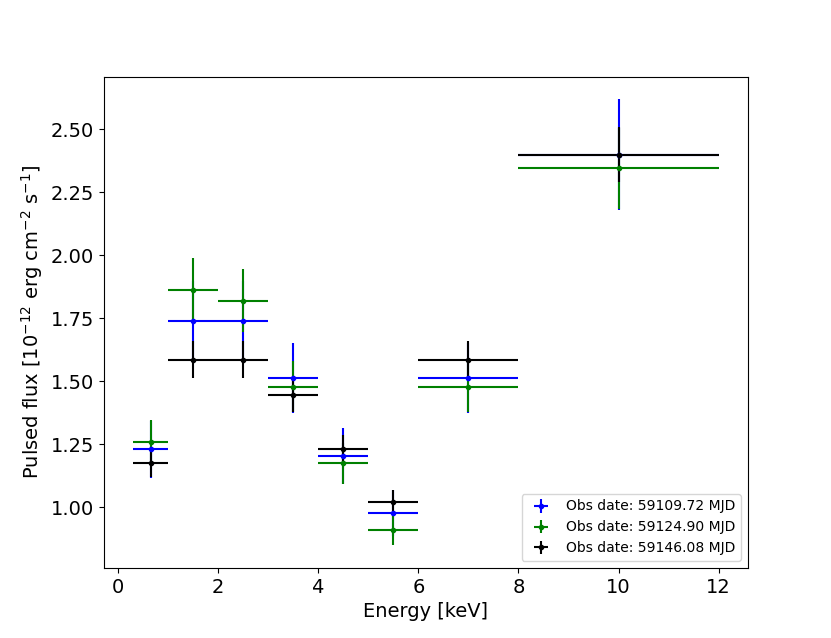}
\caption{Pulsed flux as a function of energy for the three post-outburst {\it{XMM-Newton}} observations.}
\label{Fig:pfrac_energy}
\end{figure}
%------------------------

\subsection{Pulse profile analysis}

\label{sec:pulse_profile}
Obtaining local measurements of the spin frequency and spin-down rate via GPR and {\tt{TEMPO}} (shown in Figure~\ref{Fig:main_plot}) enabled us to compute the pulse profiles by folding on the optimal ephemeris using routines in {\tt{Stingray}}\footnote{\url{https://docs.stingray.science/index.html}} \citep{huppenkothen19}. These profiles are shown in Figs.~\ref{Fig:nicer_pprof} and ~\ref{Fig:xmm_pulse_prof} for the four separate instruments. In the case of {\it{NICER}} and {\it{Swift}} the pulse profiles from several observations (indicated at the top of each figure) were combined to produce a stacked profile. The pulse profiles generally appear sinusoidal with minimal contribution from higher order harmonics $>4$. However, we note that for the period just after the 2020 outburst (i.e. MJD 59055-59085), the stacked {\it{NICER}} profile features a shoulder to the main sinusoidal component, which was originally found by \citealt{hu23}.  

We determined the long-term evolution of the pulsed flux as illustrated in Figure~\ref{Fig:main_plot} by performing phase resolved spectroscopy, which is discussed further in Section~\ref{sec:Xspec}. It is apparent that the pulsed flux increases notably on 26th June 2020 (see also \citealt{kuiper20} and \citealt{hu23}), which is coincident with a (spin-up) timing glitch and a milli-second X-ray burst that was observed a few days later via {\it{Swift}}-BAT (i.e. on MJD 59063; \citealt{kuiper20}). 

We also examined how the pulsed flux of the source varies as a function of energy for the {\it{XMM-Newton}} observations, as indicated in Figure~\ref{Fig:pfrac_energy}. It should be noted that we compute the pulsed flux as a function of energy rather than the pulsed fraction, since the total flux (in this case) does not discount the contribution from the pulsar wind nebula, thus biasing its estimates. On the other hand, the pulsed flux was estimated using phase resolved spectroscopy (see section 5), and hence subtracts the emission from the PWN. The pulsed flux appears to show curvature at energies below 6 keV, while increasing to higher values above 10 keV. This clearly implies the presence of two spectral components to the pulsed emission (i.e. a blackbody and a power-law: see section 5 for more details on the modelling of this emission). 

Finally, we illustrate how the morphology of the {\it{XMM-Newton}} pulse profiles vary with energy in Figure~\ref{Fig:xmm_pulse_prof}. As shown in the figure, we do not observe a significant change in the morphology of the pulse profiles or the occurrence of any phase shifts of the primary sinusoidal component with increasing energy. For some observations the pulse profile does appear to be more complex than a sinusoid, but no clear trends are observed as a function of energy or observation date.  

\section{Diffuse emission}

%------- FIGURE ---------
%------------------------

%------- FIGURE ---------
\begin{figure*}
\centering
\includegraphics[scale=0.6]{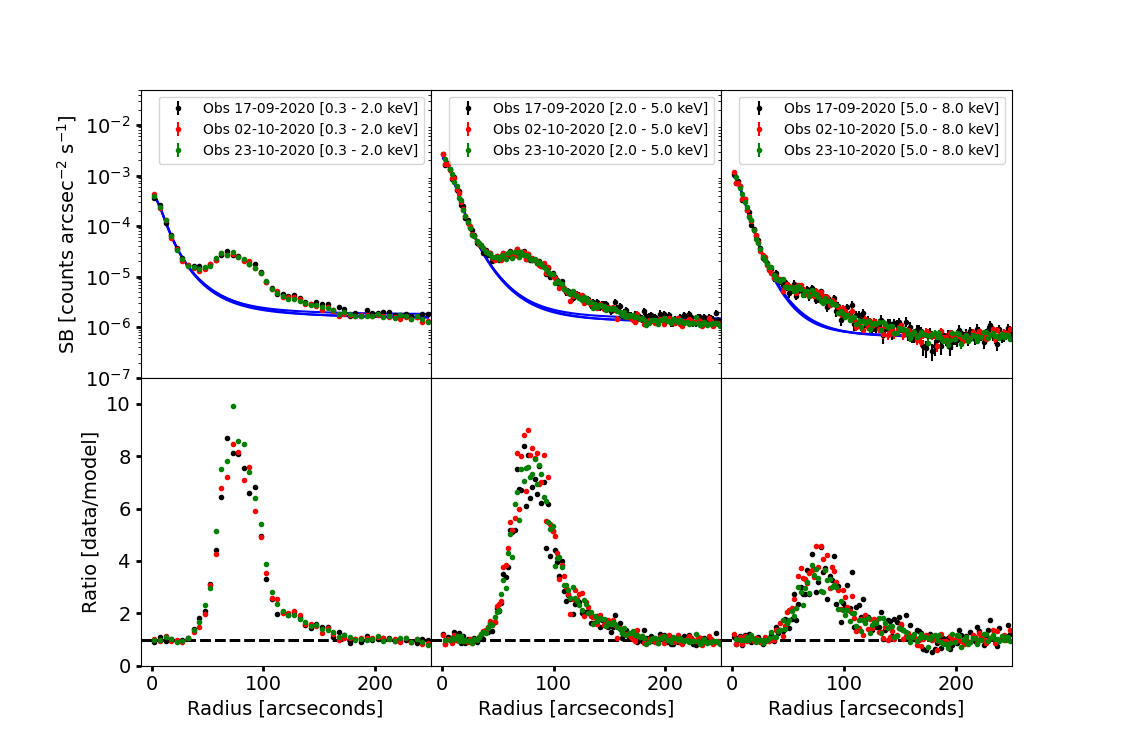}
\caption{Surface brightness (SB) profiles of the three {\xmm} observations in three different energy bands up to a radial distance of 250 arcseconds from the position of \hbpsr\,. In each case, the radial profile was fit with a King PSF model (after excluding the radial range of 35-160 arcseconds to avoid the contribution from the supernova remnant; see text; blue lines) with the addition of a constant background (black solid lines) in order to analyse the extension of the source. The residuals to the best-fit are shown in the bottom panel.}
\label{Fig:rad_profiles}
\end{figure*}
%------------------------

To analyse the extended emission surrounding the source, we extracted surface brightness profiles of the three \xmm\ observations in three different energy bands (0.3--2\,keV, 2--5\,keV and 5--8\,keV) out to a radius of 250 arcseconds (in increments of 2 arcseconds). These are shown in Figure~\ref{Fig:rad_profiles} and were fitted with an analytical King PSF model plus a constant background term $b_{0}$ such that the surface brightness $S_{b}$ was modelled as:
\begin{equation}
    S_{b} = \frac{a_0}{\big (1 + (r/r_{c})^{2}\big)^{\alpha}} + b_{0}~,
\end{equation}
where $r$ is the distance from the centre of the source, $a_{0}$, $r_{c}$, $\alpha$ and $b_{0}$ are free parameters determined by the fit. While fitting the radial profiles with this model, we excluded the radial region between 35--160 arcseconds in order to avoid the contribution from the Kes 75 supernova remnant. However, for visual purposes we overlay the emission from the remnant in Figure~\ref{Fig:rad_profiles} to illustrate the clear excess in $S_{b}$ in the above mentioned radial regions. The excess appears very strong in the 0.3--2 keV and 2--5 keV energy bands, but decreases in strength above 5 keV, illustrating the energy dependence of the emission from the remnant. As expected, there appears to be little variability in the excess emission between the three observations for all energy bands. 

%------- FIGURE ---------
\begin{figure*}
\centering
\includegraphics[scale=0.7]{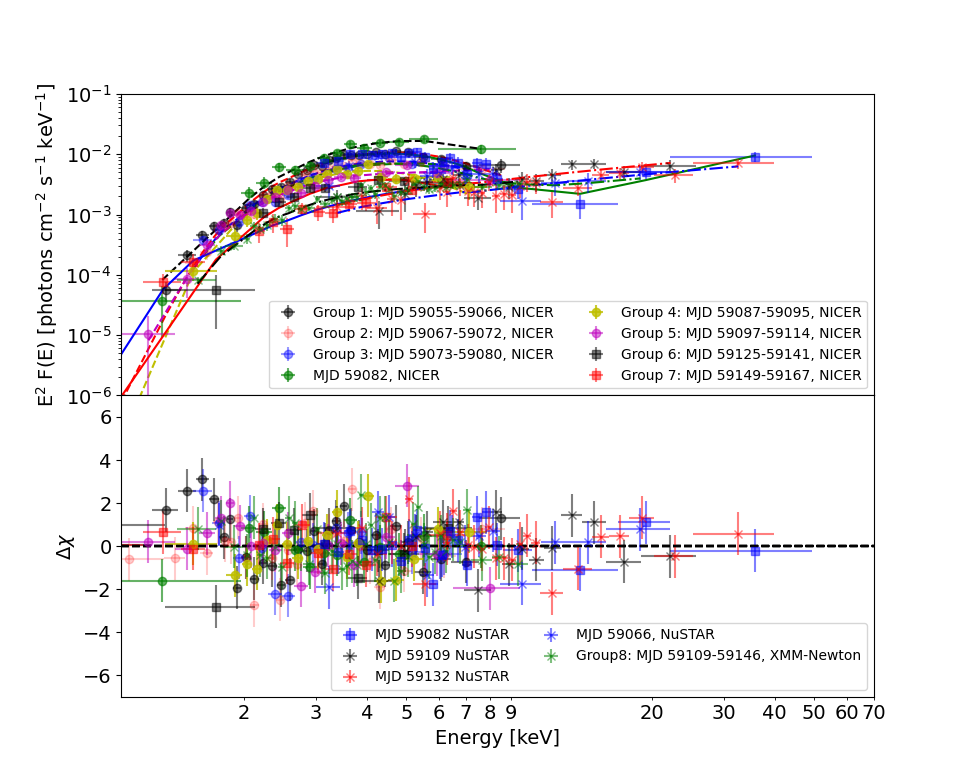}
\caption{{\it{Top panel}}: Absorbed ($N_{\text{H}} = (5.5 \pm 0.3) \times 10^{22}$ cm$^{-2}$) unfolded (deconvolved) pulsed energy spectra of \hbpsr\, in the aftermath of the 2020 outburst using {\it{NICER}}, {\it{NuSTAR}} and {\it{XMM-Newton}} observations (see text). For display purposes, the {\it{NICER}} spectra were separated into seven groups, where in each group the spectra were stacked using the {\tt{addspec}} tool (for the observation date ranges indicated in the legend). Similarly, the three {\it{XMM-Newton}} spectra were also combined. However, for spectral fitting we used the original spectra for each observation (see Table\,\ref{Tab:Table1}). Each spectrum was fit with an absorbed blackbody plus power-law model with the column density tied between the different observations. The results clearly demonstrate the gradual fading of the thermal component (below 10 keV) during the outburst, while the power-law component (above 8 keV) remains unchanged. The detailed best-fit parameters are reported in Table\,\ref{Tab:Table1}. The residuals of the best-fit model are also shown in the bottom panel of the figure.}
\label{Fig:xmmnicerspec}
\end{figure*}
%-------------------------

\section{Phase resolved X-ray spectroscopy}
\label{sec:Xspec}

We made use of the \nicer, \nustar\ and \xmm\ EPIC-pn observations to characterise the spectral properties of the pulsed emission over the energy energy range of 1-70 keV, using the {\tt{Xspec}} software package (\citealt{arnaud96}). In order to isolate the pulsed spectral component, we used the spin parameters derived from the timing analysis discussed in the previous section to perform phase-resolved spectroscopy. For the {\it{NICER}} observations, we implemented the {\tt{photonphase}} tool to assign a pulse phase to each event for a given observation based on the available timing ephemeris, and extracted the spectra for those intervals corresponding to the brightest (on) and faintest (off) quarters of the pulse cycle using {\tt{niextract-events}} and {\tt{niextspect}}. This was followed by the generation of the ancillary response (ARF) files and response matrix (RMF) files via the {\tt{nicerarf}} and {\tt{nicerrmf}} tools. We note that care was taken to ensure that the non-barycentered event files were used to generate the ARF and RMF files, since it is known that the {\nicer} pipeline could fail to produce the correct responses for barycentered event files. The pulse phase column generated on the barycentered event files were copied over to the non-barycentered files using {\tt{ftpaste}}, and appropriate phase filters were then applied to extract the pulse-on and pulse-off spectra. The {\tt{nicerarf}} tool requires the source position as input, since the detector quantum efficiency depends on the off-axis angle, and so we used the source position available from the {\it{Chandra}} observations. During the process, we also used the filter file for each observation containing key information regarding various screening criteria including passages of the spacecraft through the South Atlantic Anomaly. The pulsed spectra were finally extracted by subtracting the pulse-off spectrum from the pulse-on spectrum, and were grouped using an optimal binning algorithm from \cite{kaastra16} via the {\tt{ftgrouppha}} tool, with at-least 200 counts per energy bin. We further placed the pulsed {\it{NICER}} spectra into eight distinct groups, chosen to correspond to eight different flux levels spanning the entire duration of the outburst. Spectra in each group were combined using the tool {\tt{addspec}}.

In order to process the {\nustar} data for phase resolved spectroscopy, we generated good time interval (GTI) files corresponding to the brightest and faintest quarters of the pulse cycle. The pulsed spectra were subsequently extracted with the {\tt{nuproducts}} tool, adopting a source region size as discussed in Section~\ref{sec:Reduction}. The spectra were re-binned using {\tt{ftgrouppha}}, with at-least 50 counts per energy bin. 

For {\xmm}, the spectral extraction procedure was broadly similar. We selected source and background events according to the procedure discussed in Section~\ref{sec:Reduction}. The {\tt{photonphase}} tool was used to assign a pulse phase to each event by folding the data on the best-fit ephemeris. We then selected the quarters of the pulse cycle corresponding to the brightest and faintest emission to generate a ``pulse-on'' and a ``pulse-off'' spectrum, and subtracted the latter from the former. We generated ancillary response files and response matrix files using the {\tt{arfgen}} and {\tt{rmfgen}} tasks, respectively. All spectra were re-binned using the {\tt{ftgrouppha}} tool adopting at-least 200 counts per energy bin, and the spectra from the three instruments were read into {\tt{Xspec}}. Finally, we found that the {\swift} data were inappropriate for phase resolved analysis due to the paucity of counts, although they were used to estimate the pulsed flux.

In Figure~\ref{Fig:xmmnicerspec}, we present the pulsed spectra over the 1 - 70 keV energy range. We proceeded to fit the spectra collectively with an absorbed single temperature blackbody plus power-law model. The blackbody temperature and normalisation were allowed to vary for each spectrum, while the column density was tied between the different observations. The power-law slope and normalisation were allowed to vary only for the NuSTAR spectra, but for the {\xmm} and {\nicer} spectra, due to their limited energy range, the power-law parameters were fixed to (weighted) average values found for the four {\nustar} spectra. It should be noted however that the power-law component remains largely unchanged, as illustrated in Figure~\ref{Fig:xmmnicerspec}. In all cases, we used the abundances by \cite{wilms00}. The best-fit parameters of the spectral analysis are reported in Table\,\ref{Tab:Table1}, with an overall goodness-of-fit that is statistically acceptable (with $\chi^{2}$/dof = 1372.0/1081).  

\section{Radio observations}
\label{sec:radioobs}
PSR\,J1846--0258 was observed in radio with the Parkes-Murriyang telescope on August 5 2020. Observations, obtained through a ToO (ID PX065), were performed with the Ultra-Wide-bandwidth, Low-frequency receiver \citep[UWL, ][]{uwl} having a total bandwidth of 3328 MHz centered at 2368 MHz. The band was split into 3328 frequency channels. Full Stokes data were recorded and 8-bit sampled every 512 $\mu$s for 3.9 h, starting at UT 08:37. The data were searched for periodic pulsation and for single dispersed pulses.
 
%%%%%%%%%%%%%%% TABLE %%%%%%%%%%%%%%%%%%%%%%%%%%%%%
%\startlongtable
\begin{deluxetable*}{ccccccc}
%\centerwidetable
\centering
\tablecaption{\label{Tab:Table1}: The best-fit parameters resulting from fitting the pulsed spectra of all three instruments as discussed in the text. The column density was tied between the different observations, yielding N$_{\text{H}}$ = (5.5 $\pm$ 0.3) $\times 10^{22}$ cm$^{-2}$. All errors are quoted at the 1$\sigma$ confidence level.}
\tablecolumns{7}
%\tablenum{1}
\tablewidth{0pt}
\tablehead{
\colhead{Instrument} &
\colhead{Epoch [MJD]} &
\colhead{$kT_{\text{bb}}$ [keV]} &
\colhead{$R_{\text{bb}}$ [km]\tablenotemark{\footnotesize a}} &
\colhead{Total flux [10$^{-11}$ erg cm$^{-2}$ s$^{-1}$]\tablenotemark{\footnotesize b}} &
\colhead{BB flux [10$^{-11}$ erg cm$^{-2}$ s$^{-1}$]\tablenotemark{\footnotesize b}} &
\colhead{$\chi^{2}$/dof\tablenotemark{\footnotesize c}}\\}
\startdata
%%%%%%%%%%%%%%%%%%%%%%%%%%%%%%%%%%%%%%%%%%%%%%%%%%%%%%%%%%%%%%%%%%%%%%%%%%%%%%%%%%%%%%%%%%%%%
\nicer & 59026 & 1.0 $\pm$ 0.3 & 0.5 $\pm$ 0.3 & 1.3 $\pm$ 0.1 & 0.6 $\pm$ 0.1 & 43.4/44 \\
\nicer & 59055 & 1.1 $\pm$ 0.2 & 1.0$^{+0.5}_{-0.4}$ & 3.6 $\pm$ 0.3 & 3.1 $\pm$ 0.5 & 18.2/18 \\
\nicer & 59056 & 1.1 $\pm$ 0.2 & 0.9$^{+0.5}_{-0.4}$ & 3.4 $\pm$ 0.2 & 2.8 $\pm$ 0.4 & 30.3/19 \\
\nicer & 59064 & 0.95 $\pm$ 0.1 & 1.0 $\pm$ 0.4 & 2.6 $\pm$ 0.1 & 2.0 $\pm$ 0.2 & 64.1/38 \\
\nicer & 59067 & 1.0 $\pm$ 0.1 & 0.8 $\pm$ 0.3 & 2.5 $\pm$ 0.1 & 1.9 $\pm$ 0.3 & 49.9/30 \\
\nicer & 59068 & 1.0 $\pm$ 0.2 & 0.8$^{+0.6}_{-0.4}$ & 2.3 $\pm$ 0.2 & 1.7 $\pm$ 0.4 & 18.5/20 \\
\nicer & 59070 & 1.0 $\pm$ 0.1 & 0.8 $\pm$ 0.3 & 2.4 $\pm$ 0.1 & 1.7 $\pm$ 0.3 & 37.9/41 \\
\nicer & 59072 & 1.0 $\pm$ 0.2 & 0.8$^{+0.5}_{-0.4}$ & 2.1 $\pm$ 0.2 & 1.4 $\pm$ 0.4 & 16.5/24 \\
\nicer & 59073 & 0.9 $\pm$ 0.2 & 0.9$^{+0.4}_{-0.3}$ & 2.1 $\pm$ 0.1 & 1.5 $\pm$ 0.2 & 42.2/33 \\
\nicer & 59075 & 1.0 $\pm$ 0.2 & 0.7$^{+0.5}_{-0.3}$ & 2.0 $\pm$ 0.1 & 1.4 $\pm$ 0.3 & 24.5/22 \\
\nicer & 59079 & 0.9 $\pm$ 0.2 & 1.0$^{+0.5}_{-0.4}$ & 2.4 $\pm$ 0.2 & 1.7 $\pm$ 0.3 & 36.6/28 \\
\nicer & 59081 & 1.1 $\pm$ 0.2 & 0.7 $\pm$ 0.3 & 3.0 $\pm$ 0.1 & 2.3 $\pm$ 0.4 & 44.0/33 \\
\nicer & 59086 & 1.0 $\pm$ 0.2 & 0.8$^{+0.4}_{-0.3}$ & 2.3 $\pm$ 0.5 & 1.7 $\pm$ 0.4 & 38.5/35 \\
\nicer & 59088 & 0.9 $\pm$ 0.2 & 0.8$^{+0.4}_{-0.3}$ & 2.0 $\pm$ 0.1 & 1.3 $\pm$ 0.3  & 86.2/45 \\
\nicer & 59092 & 0.8 $\pm$ 0.2 & 0.9$^{+0.7}_{-0.5}$ & 1.6 $\pm$ 0.1 & 0.9 $\pm$ 0.3 & 68.1/41 \\
\nicer & 59098 & 1.0 $\pm$ 0.2 & 0.7$^{+0.5}_{-0.3}$ & 1.8 $\pm$ 0.1 & 1.2 $\pm$ 0.2 & 65.0/48 \\
\nicer & 59102 & 0.7 $\pm$ 0.2 & 1.2$^{+1}_{-0.8}$ & 1.4 $\pm$ 0.2 & 0.8 $\pm$ 0.3 & 46.9/40 \\
\nicer & 59122 & 1.2 $\pm$ 0.4 & 0.35$^{+0.5}_{-0.2}$ & 1.2 $\pm$ 0.1 & 0.6 $\pm$ 0.3 & 49.8/41 \\
\nicer & 59137 & 0.9 $\pm$ 0.2 & 0.5$^{+0.6}_{-0.3}$ & 1.2 $\pm$ 0.1 & 0.5 $\pm$ 0.3 & 52.5/28 \\
\nicer & 59145 & 0.6 $\pm$ 0.4 & 0.8$^{+6}_{-0.6}$ & 0.9 $\pm$ 0.2 & $<0.7$ & 66.3/60 \\
\nicer & 59154 & 0.2 $\pm$ 0.1 & $<170$ & 1.5 $\pm$ 0.2 & $<1.5$ & 92.3/49 \\
\nicer & 59165 & 0.2 $\pm$ 0.1 & $<3670$ & 1.5 $\pm$ 0.2 & $<1.6$ & 53.2/40 \\
\nicer & 59174 & $0.4^{+0.1}_{-0.3}$ & $<18000$ & 0.6 $\pm$ 0.2 & $<0.2$ & 28.8/16 \\
\xmm & 59109 & $<0.9$ & $<0.5$ & 0.9 $\pm$ 0.2 & $<0.2$ & 75.6/74 \\
\xmm & 59125 & 0.7$^{+0.3}_{-0.2}$ & 0.5$^{+1}_{-0.3}$ & 1.0 $\pm$ 0.1 & 0.3 $\pm$ 0.1 & 77.9/73 \\
\xmm & 59145 & 0.9$^{+0.6}_{-0.4}$ & 0.3$^{+0.6}_{-0.1}$ & 0.9 $\pm$ 0.1 & 0.2 $\pm$ 0.1 & 84.0/84 \\
%%%%%%%%%%%%%%%%%%%%%%%%%%%%%%%%%%%%%%%%%%%%%%%%%%%%%%%%%%%%%%%%%%%%%%%%%%%%%%%%%%%%%%%%%%%%%
\hline
\hline
%%%%%%%%%%%%%%%%%%%%%%%%%%%%%%%%%%%%%%%%%%%%%%%%%%%%%%%%%%%%%%%%%%%%%%%%%%%%%%%%%%%%%%%%%%%%%
Instrument & Epoch & $kT_{\text{bb}}$ [keV] & $R_{\text{bb}}$ [km] & Total flux [10$^{-11}$ erg cm$^{-2}$ s$^{-1}$]\tablenotemark{\footnotesize d} & Photon index & $\chi^{2}$/dof \\
&&&&&&\\
\hline
\nustar & 59066 & 0.9 $\pm$ 0.1 & 0.8$^{+1}_{-0.4}$ & 3.1 $\pm$ 0.4 & 1.0 $\pm$ 0.7 & 22.1/18 \\
\nustar & 59082 & 0.9 $\pm$ 0.1 & 1.0$^{+0.5}_{-0.3}$ & 3.5 $\pm$ 0.2 & $<1.5$ & 26.7/26 \\
\nustar & 59109 & - & - & 3.5 $\pm$ 0.1 & 1.4 $\pm$ 0.2 & 26.0/22 \\
\nustar & 59132 & - & - & 2.5 $\pm$ 0.3 & 1.3 $\pm$ 0.2 & 32.2/35\\
\hline
%%%%%%%%%%%%%%%%%%%%%%%%%%%%%%%%%%%%%%%%%%%%%%%%%%%%%%%%%%%%%%%%%%%%%%%%%%%%%%%%%%%%%%%%%%%%%
&&&&&& 1372.0/1081
\enddata
\tablecomments{
\tablenotetext{a}{Assuming a source distance of 5.8\,kpc \citep{verbiest12}.}
\tablenotetext{b}{Energy range: 1-10 keV}
\tablenotetext{c}{dof: Degrees of freedom}
\tablenotetext{d}{Energy range: 3-79 keV}}
\label{Tab:Table1}
\end{deluxetable*}

The search for a periodic signal has been performed both in the entire observational bandwidth and in three separate sub-bands ($702-1000$\,MHz, $1000-2000$\,MHz, $2000-4032$\,MHz). Data were folded with the partially coherent solution covering the radio epoch and with dispersion measures (DMs) from $0$ to $2000$\,pc\,cm$^{-3}$. Folded archives with 1-MHz frequency resolution and  512 time bins across the folded profile were created with a DM step of $200$\,pc\,cm$^{-3}$. After radio frequency interference (RFI) removal, each archive was searched over a period range spanning +/- 3.5 $\mu$s and +/- $100$\,pc\,cm$^{-3}$ around the nominal values. 

The single-pulse search was done by implementing an iterative sub-banded search \citep[similarly to ][]{kumarsub}, in addition to the search in the full observational band. The reason for sub-banding is that repeating Fast Radio Bursts (FRBs) have been observed as band-limited sources, meaning that the emission is typically limited to as little as $\sim 100$~MHz, especially at the lowest frequencies \citep{kumarsub}, and this would result in  a large loss of sensitivity (or of the burst altogether) in the full-band search. In the sub-banded search, we searched for bursts by inspecting iteratively smaller bandwidths of widths $[2 \times 1664, 4 \times 832, 8 \times 416, 16 \times 208, 32 \times 104]$\,MHz. For each sub-band, we also considered overlapping adjacent sub-bands (by shifting the bands by half widths) in order not to miss events between two adjoining sub-bands. This results to a total of $[1 \times 3328, 3 \times 1664, 7 \times 832, 15 \times 416, 31 \times 208, 63 \times 104]$\,MHz bands processed. Data were dedispersed from $0$ to $2000$\,pc\,cm$^{-3}$ and we searched for excesses by convolving the frequency-collapsed time series with a top-hat function with trial widths logarithmically spaced and ranged up to a maximum width of 512 bins. 

No clear sign of either periodic radio emission or sporadic single pulses was found above a signal-to-noise ratio of 8 and 10, respectively (see also the works by \citealt{blum21} at 2 GHz, \citealt{mickaliger2020} at 1.5 GHz and \citealt{majid20} at 8.3 GHz and 13.9 GHz). From the single pulse search we redetected the RRAT\,J1846-0257 \citep{rratredetected}, whose position is $\sim 1.3$\,arcmin from the nominal center of the telescope beam pointing, with FWHM beam-width of $\sim 4.9$\,arcmin at $4032$\,MHz. By exploiting the standard radiometer equation for folded pulsar profiles and single pulses \citep{handbookpulsar, cordes_2003_sp}, we can set our upper limits both for radio periodic emission and single bursts. We assume for the UWL a telescope gain of $G = 1.8$\,Jy\,K$^{-1}$ and an average system temperature without the sky contribution of $\sim 22$\,K \citep[see Fig.\,4 from][]{uwl}. The frequency-dependent sky temperatures at the source pointing were computed assuming a pivotal temperature of $342.3$\,K at $408$\,MHz \citep{haslam} and scaled with a spectral index of $-2.7$.

\begin{table}[h]
\label{Tab:Table2}
\caption{Radio emission upper limits (quoted at the 1$\sigma$ confidence level). For the periodic emission we report the peak flux density $S_{\nu}$ upper limit, whereas for single pulses we report the fluence density $F_{\nu}$ (see Sec.\,\ref{sec:radioobs} for more details).}
\centering
\begin{tabular}{llc}
\hline
\hline
Periodic Emission & Frequency Band & $S_{\nu}$ \\ 
                  & (MHz)          &  (mJy) \\ 
\hline 
& 702-4032  & 0.01  \\ 
& 702-1000  & 0.03  \\ 
& 1000-2000 & 0.02 \\ 
& 2000-4032 & 0.01 \\ 
\hline
\hline 
Single Pulses & Frequency Band & $F_{\nu}$ \\ 
              & (MHz) &  (Jy ms) \\ 
\hline 
& 702-4032  & 0.2  \\
& 702-2366  & 0.3  \\ 
& 702-1534  & 0.9  \\ 
& 702-1118  & 2  \\ 
& 702-954   & 3  \\ 
& 702-806   & 4  \\
\hline                  
\end{tabular}
\label{Tab:Table2}
\end{table}
%%% -------------------------------------------

For the putative periodic radio emission, assuming a fiducial period of $P = 0.326$\,s and a duty cycle of 10\,\%, we report the peak flux density upper limits in the four bands processed in Table~\ref{Tab:Table2}.

Table~\ref{Tab:Table2} also reports our fluence density upper limits for the emission of single pulses in the full-band and the sub-bands (the upper limits are quoted at the 1$\sigma$ confidence level). Here we assumed a pulse/burst with 1\,ms duration and considered, for each chosen bandwidth, the lowest frequency sub-band (as this is the one with the highest sky temperature). The limits in the sub-banded search are, as expected, higher than the limit in the full band. This limit, however, works under the assumption that the burst is visible within the full bandwidth and that it has flat spectrum. In the hypothesis that the burst is band-limited, the sub-banded search can become more sensitive as the average signal-to-noise will be less affected by the noise contribution (for details see Trudu et al., in prep.).

\newpage
\section{Discussion}
\label{sec:discussion}

%------- FIGURE ---------
\begin{figure*}
\vspace{0.5cm}
\begin{center}
\includegraphics[scale=0.8]{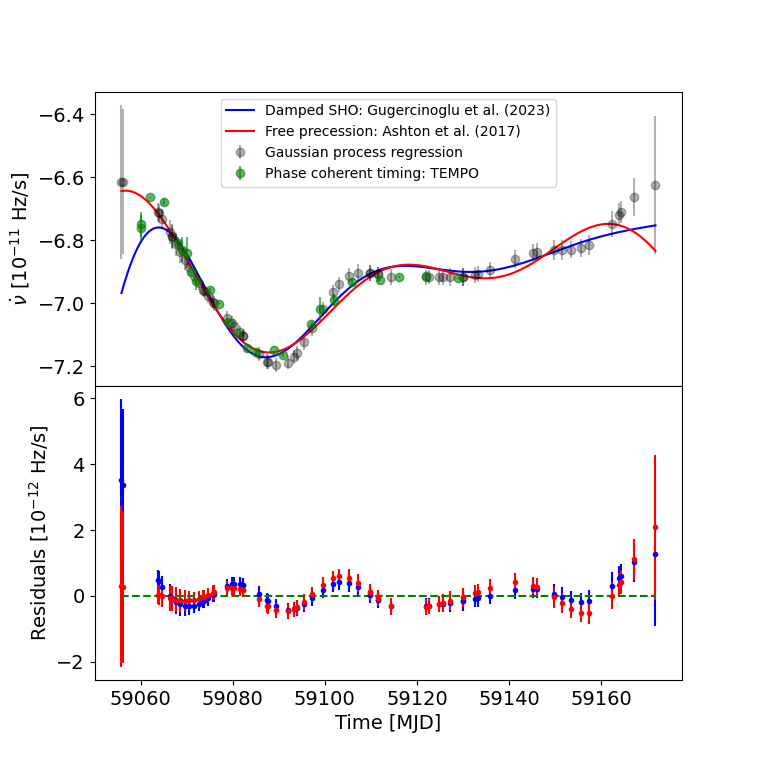}
\end{center}
\caption{The evolution in the frequency derivative $\dot{\nu}$ modelled with the combination of spin-down terms and damped harmonic oscillator (blue curve; \citealt{gugercinoglu23}) and free precession (red curve; \citealt{ashton17}), as indicated in the top panel. The residuals (data - model) are shown in the bottom panel, which clearly demonstrate significant scatter (indicating that these models are likely unsuitable to describe the full complex $\dot{\nu}$ evolution). To obtain a statistically acceptable fit, we fit the models to a smaller section of the data (i.e. MJD 59055 - 59135). We find for the damped SHO, $\chi^{2}$/dof = 51.5/46, while for the free precession model, $\chi^{2}$/dof = 50.5/45.}
\label{Fig:nudotmodel}
\end{figure*}
%----------------------

The source \hbpsr\ is one of the best examples of a highly energetic pulsar showing both rotation and magnetic powered emission during its bursting period. This pulsar, together with \radiohb\, and the low-field magnetars suggests how different pulsar classes can be unified based on their age, magnetic field strength and geometric configuration at birth. In this work, we studied in detail the 2020 outburst of \hbpsr. Among other findings, we found indications for an intriguing oscillation of the spin frequency derivative in the immediate aftermath of the 2020 outburst (see Figs.~\ref{Fig:main_plot} and ~\ref{Fig:spin_par_comp}). This is rather unexpected as quasi-periodic oscillations in the timing residuals of isolated pulsars are the result of timing noise that occurs on timescales of several years (see e.g. \citealt{hobbs09}). However, in this case, the timescales are significantly shorter. 

Considering a time variable coupling of the superfluid NS core and crust (i.e. the model by \citealt{gugercinoglu23}), we attempt to constrain the timescale of the quasi-periodicity by modelling the evolution of $\dot{\nu} = \dot{\Omega}/2\pi$ with a combination of a damped simple harmonic oscillator (SHO) and regular spin-down, in line with the following expression:

\begin{multline}
\dot{\Omega}(t) = 2\pi \dot{\nu}_{\text{0}} + 2\pi \ddot{\nu}_{\text{0}}t + \\
	A\Omega_{0} \exp \bigg (-\frac{t}{2 \tau } \bigg) \cos \big[\Omega_{0}t + \phi\big] -\\ 
	\frac{A}{2 \tau }\exp \bigg (-\frac{t}{2 \tau} \bigg ) \sin \big[\Omega_{0}t + \phi\big] ~\text{with}\\
	\Omega_{0} = \omega_{0} \bigg[1 - \bigg(\frac{1}{2\tau\omega_{0}}\bigg)^{2}\bigg]^{1/2}
\end{multline}

where the symbols are defined in Table~\ref{Tab:Table3} and have the same meanings as in \cite{gugercinoglu23}.
\newline
\newline
We also consider the precession model of \cite{ashton17} (triggered plausibly due to magnetic re-adjustments in the NS crust at the outburst onset, leading to deformation of the crust) in order to fit the periodic trend in the spin-down rate. We allow the precession angle to vary secularly with time (see equations 2-6 and equation 10 of \citealt{ashton17} for full details on the model).
\newline
\newline
The above two models were fit to the data using least-squares minimisation via the {\tt{kmpfit}} package (\citealt{KapteynPackage}), but the resulting fits (after convergence) were found to be statistically unacceptable. This is apparent from the residuals to the best-fit models that clearly show considerable scatter (see bottom panel of Figure~\ref{Fig:nudotmodel}). Moreover, performing a runs-test rejects the two models at a significance of above 5 sigma (with p-values below $10^{-7}$). In order to improve the goodness-of-fit and extract parameter uncertainties, we fit the models to a smaller section of the data (i.e. MJD 59055 - 59135). We show the corresponding best-fit parameters in Table~\ref{Tab:Table3} and Table~\ref{Tab:Table4}. We note that while the above models do not formally describe the full dataset, they provide a crude estimate of the oscillation period $P_{\text{osc}} = \frac{2\pi}{\Omega_{0}} \sim 50-60$ days and the damping timescale $\tau \sim 20$\,days (in the context of the damped harmonic oscillator) and the ellipticity $\epsilon$ of the NS (for the precession model), which we discuss below. We suggest leaving a detailed modelling of the spin-down rate in a future work, while noting the main points below.
\newline
\newline
In the context of the damped SHO model, the oscillation period of $\sim 50-60$ days could be consistent with Ekman oscillations (due to non-spherical geometry of the superfluid vortex lines) with $T_{\text{Ekman}} = \frac{R}{(\kappa \Omega)^{1/2}}$, where $R$ is the NS radius, $\Omega$ is the angular rotational velocity and $\kappa = 2 \times 10^{-3}$ cm$^{2}$ s$^{-1}$ is the quantised vorticity attached to each vortex line (see section 3 of \citealt{gugercinoglu23}). Moreover, the damping timescale $\tau$ of the spin-down rate is related to the linear creep timescale of the superfluid vortices, which depends on the micro-physical properties of the NS crust (see equation 5 of \citealt{gugercinoglu23}). The observed $\tau \sim 20$ days seems plausible if the surface temperature of the NS is $\sim$ 0.015 keV. We note that this is much smaller than the blackbody temperature observed during late stages of the outburst in this work (although a more precise measurement of the surface temperature would require optical and UV observations and precise constraints on the hydrogen column density). In computing the expected damping timescale, we assume typical parameter values for other micro-physical quantities of the NS (e.g. the age dependent critical angular velocity between normal matter and superfluid crust $\omega_{\text{cr}} \sim 5 \times 10^{-3}$ rad/s, the pinning energy between superfluid vortices and lattice nuclei $E_{\text{P}} \sim 10$ keV and microscopic vortex velocity around nuclei $v_{0} \sim 10^7$ cm/s; see \citealt{alpar89}; \citealt{haskell15}). 
\newline
\newline
In the context of the precession model, the magnetic ellipticity of the NS is given by $\epsilon_B = \frac{B^2R^4}{GM^2}$ and is about $5.7\times 10^{-8}$ for this source assuming that the surface dipolar magnetic field $B$ ($\sim 5 \times 10^{13}$ G) can be estimated in the standard manner via the observed spin period and spin-down-rate at outburst onset for a canonical mass of 1.4 $M_{\sun}$ and radius $R$ of 10--12 km. The estimated ellipticity seems consistent with the ellipticity inferred by modelling the spin-down rate variability (see Table~\ref{Tab:Table4}). However, we note that the $\dot{\nu}$ modulation does not appear to be correlated with (periodic) changes in the source flux, which could disfavor an interpretation in terms of precession (although this statement may depend on the size of the precession angle).

%------- TABLE ----
%\startlongtable
\begin{deluxetable}{ccccc}
%\centerwidetable
\centering
\tablecaption{The best-fit parameters for modelling the evolution of the frequency derivative $\dot{\nu}$ with a damped harmonic oscillator (\citealt{gugercinoglu23})}
\label{Tab:Table3}
\tablecolumns{2}
%\tablenum{1}
\tablewidth{0pt}
\tablehead{
\colhead{Parameter} &
\colhead{Best-fit value} \\
\colhead{} & 
\colhead{} & 
}
\startdata
Validity period [MJD] & 59055 - 59135 \\
$\phi$ [reference phase] & $-0.7 \pm 0.1$ \\
$\omega_{0}$ & 0.120 $\pm 0.002$ \\
Damping timescale, $\tau$ [days] & 20 $\pm 2$ \\
Amplitude, $A$ [rad s$^{-1}$] & (2.7 $\pm 0.2$) $\times 10^{-6}$\\
$\dot{\nu}_{\text{0}}$ [Hz s$^{-1}$] & -(7.06 $\pm$ 0.01) $\times 10^{-11}$  \\
$\ddot{\nu}_{\text{0}}$ [Hz s$^{-2}$] & (1.9 $\pm$ 0.2) $\times 10^{-19}$  \\
Oscillation frequency, $\Omega_{0}$ [rad s$^{-1}$] & (1.35 $\pm 0.03) \times 10^{-6}$ \\
\hline
$\chi^{2}$/dof & 51.5/46
\enddata
\end{deluxetable}

%------- TABLE ----
%\startlongtable
\begin{deluxetable}{ccccc}
%\centerwidetable
\centering
\tablecaption{The best-fit parameters for modelling the evolution of the frequency derivative $\dot{\nu}$ with free precession (\citealt{ashton17}).}
\label{Tab:Table4}
\tablecolumns{2}
%\tablenum{1}
\tablewidth{0pt}
\tablehead{
\colhead{Parameter} &
\colhead{Best-fit value} \\
\colhead{} & 
\colhead{} & 
}
\startdata
Validity period [MJD] & 59055 - 59135 \\
$T_{\text{ref}}$ [days after outburst] & 59076.9 $\pm 0.3$ \\
Wobble angle $\theta_{0}$ [rad] & (7.3 $\pm 0.1) \times 10^{-5}$ \\
$\dot{\theta_{0}}$ [rad s$^{-1}$] & $-(2.0 \pm 0.1) \times 10^{-6}$ \\
$\chi$ [radians] & $9.4285 \pm 0.00001$\\
$\epsilon$ & (6.0 $\pm 0.2$) $\times 10^{-8}$\\
$\dot{\nu}_{\text{0}}$ [Hz s$^{-1}$] & -(7.048 $\pm$ 0.005) $\times 10^{-11}$  \\
$\ddot{\nu}_{\text{0}}$ [Hz s$^{-2}$] & (5.2 $\pm$ 0.3) $\times 10^{-19}$  \\
$\psi_{0}$ & 0\\
\hline
$\chi^{2}$/dof & 50.5/45
\enddata
\end{deluxetable}

%------------------------

Furthermore, the magnetospheric twist model developed by \cite{beloborodov09} is often invoked to explain the spin-down and radiative properties of a large number of magnetars. In this model, the crust of the NS is prone to azimuthal displacements resulting from star-quakes, which is expected to implant twists in the current carrying bundle of poloidal field lines in the regions above the crust. The twist inflates the poloidal field lines, leading initially to a stronger dipole torque on the NS and thus increasing the spin-down rate following the outburst. However, these twists are gradually expected to decay due to the generation of screening currents (e.g. electron-positron pairs) in the magnetosphere which oppose the currents maintaining the twist. This then implies that the spin-down torque acting on the NS decreases, and hence the spin-down rate also decreases. Qualitatively, the observed evolution of $\dot{\nu}$ shown in Fig.~\ref{Fig:main_plot} could be consistent with the above model, although the presence of an oscillation in $\dot{\nu}$ during the outburst may be difficult to explain, and there is also the question of whether the timescale over which the $\dot{\nu}$ varies can be reasonably accounted for.  

We also carried out radio observations of the source with the Murriyang Parkes telescope, setting upper limits for periodic radio emission and single pulses at frequencies centered at 2368 MHz, with a bandwidth of 3328 MHz. The radio-quiet nature of the source has previously been found at other frequencies by \citealt{blum21}, \citealt{mickaliger2020} and \citealt{majid20}.

\begin{figure*}
\vspace{0.5cm}
\begin{center}
\includegraphics[scale=0.5]{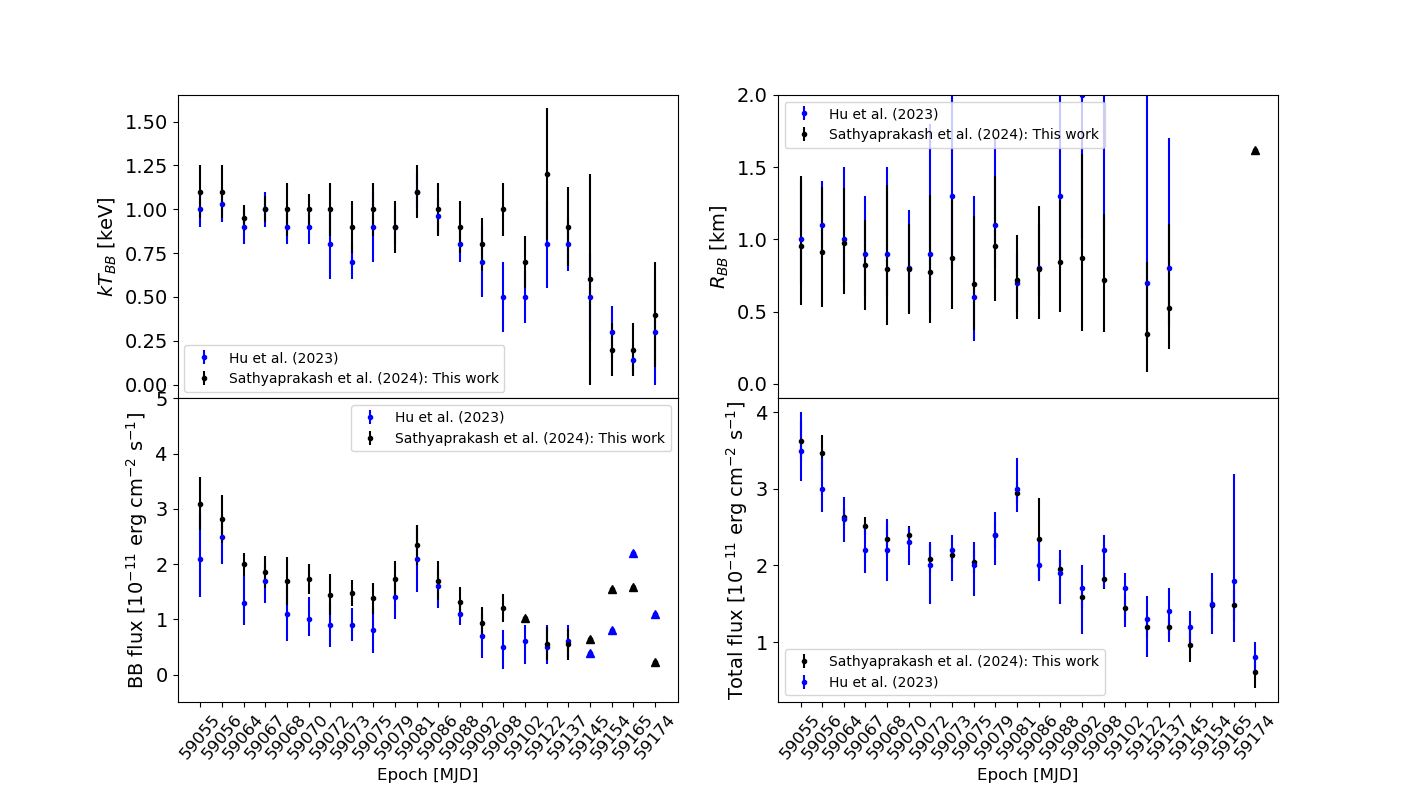}
\end{center}
\caption{The evolution in the best-fit blackbody temperatures and radii (top panels), the pulsed flux and total flux (bottom panels) during the 2020 outburst of PSR J1846-0258. The values derived in this study (black points) are compared with those derived in \citealt{hu23} (blue points). The upper limits (quoted at the 1$\sigma$ confidence level) are indicated by the arrows.}
\label{Fig:specparams}
\end{figure*}

Moreover, phase-resolved X-ray spectroscopy carried out using all four instruments clearly demonstrates that the pulsed flux of the source (computed over the 1--12\,keV energy range) begins to rise from the quiescent level at around MJD 59026 (see Fig.~\ref{Fig:main_plot}), peaks at MJD 59056, and thereafter decreases exponentially before reaching pre-outburst levels 100 days later. Moreover, the source appears to undergo a second mini-outburst (\citealt{hu23}) episode (lasting less than a day) at around MJD 59082, when the decaying trend in the pulsed flux is interrupted by a sudden increase, as suggested by the {\nicer} (ObsID: 3033290116) and {\nustar} (ObsID: 80602315004) observations.    
\newline
\newline
Our spectral analysis indicates that there exist at-least two distinct components contributing to the pulsed spectrum (over the 1.0 - 70 keV energy range): a single-temperature blackbody and a power-law, both of which are absorbed by intervening hydrogen gas along the line-of-sight. Careful spectral modelling implies the gradual fading of the thermal component during the outburst while the power-law remains largely unchanged, which was also noted recently by \cite{hu23}. We illustrate the evolution in the best-fit blackbody temperatures and radii during the outburst of the source in Figure~\ref{Fig:specparams}, comparing these values with that derived in \citealt{hu23}, finding good agreement between the two studies in general. The temperature of the thermal component seems to correlate with the blackbody (and total) flux, showing an enhancement around MJD 59082, while gradually decaying to pre-outburst values a couple of months later. The derived blackbody radius appears constant for large durations of the outburst between 0.8 - 1.2 km, before decreasing to $\sim 0.3-0.5$ km at late stages of the outburst (i.e. after MJD 59122).

%------- FIGURE ---------
%----------------------

The gradual fading of the thermal component could qualitatively be explained by the model of \cite{beloborodov09}, due to the dissipation of the twisted bundle of field lines in the magnetosphere, such that a smaller number of charged particles impact upon the NS surface. When the source returns to quiescence, the power-law begins to dominate the pulsed emission (see e.g. \citealt{kuiper18} and \citealt{gotthelf21} for an analysis over a broader energy range). 

Moreover, we estimate that the total energy released during the outburst (assuming an exponential model for the flux decay, isotropic emission in the 1 - 12 keV energy band and a source distance of 5.8 kpc) is $\sim 1.4 \times 10^{42}$ ergs. This is within a factor of three of the same quantity found for the previous outburst (in the 2 - 10 keV band) by \citet{gavriil08}. In comparison with other magnetar outbursts (see Fig.\,6 in \citealt{cotizelati18}) \hbpsr\ does not show any dissimilarities with canonical magnetars despite its hybrid nature.

\section{Acknowledgments}
RS, NR, and AI are supported by the H2020 ERC Consolidator Grant “MAGNESIA” under grant agreement No. 817661. FCZ is supported by a Ram\'on y Cajal fellowship (grant agreement RYC2021-030888-I). RS also acknowledges financial support from the Italian Ministry for University and Research, through the grants 2022Y2T94C (SEAWIND) and from INAF through LG 2023 BLOSSOM. This work was also partially supported by the program Unidad de Excelencia Mar\'ia de Maeztu CEX2020-001058-M. We thank Emilie Parent for useful discussions on the timing analysis. We also acknowledge partial support from grant SGR2021-01269 (PI: Graber/Rea). A.I.’s work has been carried out within the framework of the doctoral program in Physics at the Universitat Autonoma de Barcelona. Part of the research activities described in this paper were carried out with contribution of the NextGenerationEU funds within the National Recovery and Resilience Plan (PNRR), Mission 4 - Education and Research, Component 2 - From Research to Business (M4C2), Investment Line 3.1 - Strengthening and creation of Research Infrastructures, Project IR0000026 – Next Generation Croce del Nord. The Parkes radio telescope is part of the Australia Telescope National Facility (https://ror.org/05qajvd42) which is funded by the Australian Government for operation as a National Facility managed by CSIRO. We acknowledge the Wiradjuri people as the Traditional Owners of the Observatory site. This work has made use of the {\tt{matplotlib}} (\citealt{hunter07}), {\tt{numpy}} (\citealt{harris20}) and {\tt{scipy}} (\citealt{scipy2020}) libraries in {\tt{python}}v3.8. 

\bibliographystyle{aasjournal}
\bibliography{biblio}

\begin{thebibliography}{}
\expandafter\ifx\csname natexlab\endcsname\relax\def\natexlab#1{#1}\fi
\providecommand{\url}[1]{\href{#1}{#1}}
\providecommand{\dodoi}[1]{doi:~\href{http://doi.org/#1}{\nolinkurl{#1}}}
\providecommand{\doeprint}[1]{\href{http://ascl.net/#1}{\nolinkurl{http://ascl.net/#1}}}
\providecommand{\doarXiv}[1]{\href{https://arxiv.org/abs/#1}{\nolinkurl{https://arxiv.org/abs/#1}}}

\bibitem[{{Alpar} {et~al.}(1989){Alpar}, {Cheng}, \& {Pines}}]{alpar89}
{Alpar}, M.~A., {Cheng}, K.~S., \& {Pines}, D. 1989, \apj, 346, 823,
  \dodoi{10.1086/168063}

\bibitem[{{Archibald} {et~al.}(2015){Archibald}, {Kaspi}, {Beardmore},
  {Gehrels}, \& {Kennea}}]{archibald15a}
{Archibald}, R.~F., {Kaspi}, V.~M., {Beardmore}, A.~P., {Gehrels}, N., \&
  {Kennea}, J.~A. 2015, \apj, 810, 67, \dodoi{10.1088/0004-637X/810/1/67}

\bibitem[{{Archibald} {et~al.}(2016){Archibald}, {Kaspi}, {Tendulkar}, \&
  {Scholz}}]{archibald16}
{Archibald}, R.~F., {Kaspi}, V.~M., {Tendulkar}, S.~P., \& {Scholz}, P. 2016,
  \apjl, 829, L21

\bibitem[{{Arnaud}(1996)}]{arnaud96}
{Arnaud}, K.~A. 1996, in Astronomical Data Analysis Software and Systems V,
  Vol. 101, XSPEC: The First Ten Years, ed. G.~H. {Jacoby} \& J.~{Barnes} (ASP,
  San Francisco), 17--20

\bibitem[{{Ashton} {et~al.}(2017){Ashton}, {Jones}, \& {Prix}}]{ashton17}
{Ashton}, G., {Jones}, D.~I., \& {Prix}, R. 2017, \mnras, 467, 164,
  \dodoi{10.1093/mnras/stx060}

\bibitem[{{Beloborodov}(2009)}]{beloborodov09}
{Beloborodov}, A.~M. 2009, \apj, 703, 1044,
  \dodoi{10.1088/0004-637X/703/1/1044}

\bibitem[{{Blumer} {et~al.}(2021){Blumer}, {Safi-Harb}, {McLaughlin}, \&
  {Fiore}}]{blum21}
{Blumer}, H., {Safi-Harb}, S., {McLaughlin}, M.~A., \& {Fiore}, W. 2021, \apjl,
  911, L6, \dodoi{10.3847/2041-8213/abf11d}

\bibitem[{{Brook} {et~al.}(2016){Brook}, {Karastergiou}, {Johnston}, {Kerr},
  {Shannon}, \& {Roberts}}]{brook16}
{Brook}, P.~R., {Karastergiou}, A., {Johnston}, S., {et~al.} 2016, \mnras, 456,
  1374, \dodoi{10.1093/mnras/stv2715}

\bibitem[{{Camilo} {et~al.}(2006){Camilo}, {Ransom}, {Halpern}, {Reynolds},
  {Helfand}, {Zimmerman}, \& {Sarkissian}}]{camilo06}
{Camilo}, F., {Ransom}, S.~M., {Halpern}, J.~P., {et~al.} 2006, \nat, 442, 892,
  \dodoi{10.1038/nature04986}

\bibitem[{{Cordes} \& {McLaughlin}(2003)}]{cordes_2003_sp}
{Cordes}, J.~M., \& {McLaughlin}, M.~A. 2003, \apj, 596, 1142,
  \dodoi{10.1086/378231}

\bibitem[{{Coti Zelati} {et~al.}(2018){Coti Zelati}, {Rea}, {Pons}, {Campana},
  \& {Esposito}}]{cotizelati18}
{Coti Zelati}, F., {Rea}, N., {Pons}, J.~A., {Campana}, S., \& {Esposito}, P.
  2018, \mnras, 474, 961

\bibitem[{Esposito {et~al.}(2021)Esposito, Rea, \& Israel}]{esposito21}
Esposito, P., Rea, N., \& Israel, G.~L. 2021, Magnetars: A Short Review and
  Some Sparse Considerations, ed. T.~M. Belloni, M.~M{\'e}ndez, \& C.~Zhang
  (Berlin, Heidelberg: Springer Berlin Heidelberg), 97--142,
  \dodoi{10.1007/978-3-662-62110-3_3}

\bibitem[{{Freire} \& {Ridolfi}(2018)}]{fr18}
{Freire}, P. C.~C., \& {Ridolfi}, A. 2018, \mnras, 476, 4794,
  \dodoi{10.1093/mnras/sty524}

\bibitem[{{Gavriil} {et~al.}(2008){Gavriil}, {Gonzalez}, {Gotthelf}, {Kaspi},
  {Livingstone}, \& {Woods}}]{gavriil08}
{Gavriil}, F.~P., {Gonzalez}, M.~E., {Gotthelf}, E.~V., {et~al.} 2008, Science,
  319, 1802

\bibitem[{{Gotthelf} {et~al.}(2021){Gotthelf}, {Safi-Harb}, {Straal}, \&
  {Gelfand}}]{gotthelf21}
{Gotthelf}, E.~V., {Safi-Harb}, S., {Straal}, S.~M., \& {Gelfand}, J.~D. 2021,
  \apj, 908, 212, \dodoi{10.3847/1538-4357/abd32b}

\bibitem[{{Gotthelf} {et~al.}(2000){Gotthelf}, {Vasisht}, {Boylan-Kolchin}, \&
  {Torii}}]{gotthelf20}
{Gotthelf}, E.~V., {Vasisht}, G., {Boylan-Kolchin}, M., \& {Torii}, K. 2000,
  \apjl, 542, L37, \dodoi{10.1086/312923}

\bibitem[{{G{\"u}gercino{\u{g}}lu} {et~al.}(2023){G{\"u}gercino{\u{g}}lu},
  {K{\"o}ksal}, \& {G{\"u}ver}}]{gugercinoglu23}
{G{\"u}gercino{\u{g}}lu}, E., {K{\"o}ksal}, E., \& {G{\"u}ver}, T. 2023,
  \mnras, 518, 5734, \dodoi{10.1093/mnras/stac3516}

\bibitem[{Harris {et~al.}(2020)Harris, Millman, van~der Walt, Gommers,
  Virtanen, Cournapeau, Wieser, Taylor, Berg, Smith, Kern, Picus, Hoyer, van
  Kerkwijk, Brett, Haldane, del R{'{\i}}o, Wiebe, Peterson,
  G{'{e}}rard-Marchant, Sheppard, Reddy, Weckesser, Abbasi, Gohlke, \&
  Oliphant}]{harris20}
Harris, C.~R., Millman, K.~J., van~der Walt, S.~J., {et~al.} 2020, Nature, 585,
  357, \dodoi{10.1038/s41586-020-2649-2}

\bibitem[{{Haskell} \& {Melatos}(2015)}]{haskell15}
{Haskell}, B., \& {Melatos}, A. 2015, International Journal of Modern Physics
  D, 24, 1530008, \dodoi{10.1142/S0218271815300086}

\bibitem[{{Haslam} {et~al.}(1982){Haslam}, {Salter}, {Stoffel}, \&
  {Wilson}}]{haslam}
{Haslam}, C.~G.~T., {Salter}, C.~J., {Stoffel}, H., \& {Wilson}, W.~E. 1982,
  \aaps, 47, 1

\bibitem[{{Helfand} {et~al.}(2003){Helfand}, {Collins}, \&
  {Gotthelf}}]{helfand03}
{Helfand}, D.~J., {Collins}, B.~F., \& {Gotthelf}, E.~V. 2003, \apj, 582, 783,
  \dodoi{10.1086/344725}

\bibitem[{{Hobbs}(2009)}]{hobbs09}
{Hobbs}, G. 2009, arXiv e-prints, arXiv:0911.5534,
  \dodoi{10.48550/arXiv.0911.5534}

\bibitem[{{Hobbs} {et~al.}(2010){Hobbs}, {Lyne}, \& {Kramer}}]{hobbs10}
{Hobbs}, G., {Lyne}, A.~G., \& {Kramer}, M. 2010, \mnras, 402, 1027,
  \dodoi{10.1111/j.1365-2966.2009.15938.x}

\bibitem[{{Hobbs} {et~al.}(2020){Hobbs}, {Manchester}, {Dunning}, {Jameson},
  {Roberts}, {George}, {Green}, {Tuthill}, {Toomey}, {Kaczmarek}, {Mader},
  {Marquarding}, {Ahmed}, {Amy}, {Bailes}, {Beresford}, {Bhat}, {Bock},
  {Bourne}, {Bowen}, {Brothers}, {Cameron}, {Carretti}, {Carter}, {Castillo},
  {Chekkala}, {Cheng}, {Chung}, {Craig}, {Dai}, {Dawson}, {Dempsey}, {Doherty},
  {Dong}, {Edwards}, {Ergesh}, {Gao}, {Han}, {Hayman}, {Indermuehle},
  {Jeganathan}, {Johnston}, {Kanoniuk}, {Kesteven}, {Kramer}, {Leach},
  {Mcintyre}, {Moss}, {Os{\l}owski}, {Phillips}, {Pope}, {Preisig}, {Price},
  {Reeves}, {Reilly}, {Reynolds}, {Robishaw}, {Roush}, {Ruckley}, {Sadler},
  {Sarkissian}, {Severs}, {Shannon}, {Smart}, {Smith}, {Smith}, {Sobey},
  {Staveley-Smith}, {Tzioumis}, {van Straten}, {Wang}, {Wen}, \&
  {Whiting}}]{uwl}
{Hobbs}, G., {Manchester}, R.~N., {Dunning}, A., {et~al.} 2020, \pasa, 37,
  e012, \dodoi{10.1017/pasa.2020.2}

\bibitem[{{Hu} {et~al.}(2023){Hu}, {Kuiper}, {Harding}, {Younes}, {Blumer},
  {Ho}, {Enoto}, {Espinoza}, \& {Gendreau}}]{hu23}
{Hu}, C.-P., {Kuiper}, L., {Harding}, A.~K., {et~al.} 2023, \apj, 952, 120,
  \dodoi{10.3847/1538-4357/acd850}

\bibitem[{Hunter(2007)}]{hunter07}
Hunter, J.~D. 2007, Computing in Science \& Engineering, 9, 90,
  \dodoi{10.1109/MCSE.2007.55}

\bibitem[{{Huppenkothen} {et~al.}(2019){Huppenkothen}, {Bachetti}, {Stevens},
  {Migliari}, {Balm}, {Hammad}, {Khan}, {Mishra}, {Rashid}, {Sharma}, {Martinez
  Ribeiro}, \& {Valles Blanco}}]{huppenkothen19}
{Huppenkothen}, D., {Bachetti}, M., {Stevens}, A.~L., {et~al.} 2019, \apj, 881,
  39, \dodoi{10.3847/1538-4357/ab258d}

\bibitem[{{Kaastra} \& {Bleeker}(2016)}]{kaastra16}
{Kaastra}, J.~S., \& {Bleeker}, J.~A.~M. 2016, \aap, 587, A151,
  \dodoi{10.1051/0004-6361/201527395}

\bibitem[{{Kaspi} \& {Beloborodov}(2017)}]{kaspi17}
{Kaspi}, V.~M., \& {Beloborodov}, A.~M. 2017, \araa, 55, 261

\bibitem[{{Krimm} {et~al.}(2020){Krimm}, {Lien}, {Page}, {Palmer}, \&
  {Tohuvavohu}}]{krimm20}
{Krimm}, H.~A., {Lien}, A.~Y., {Page}, K.~L., {Palmer}, D.~M., \& {Tohuvavohu},
  A. 2020, The Astronomer's Telegram, 13913, 1

\bibitem[{{Kuiper} {et~al.}(2020){Kuiper}, {Harding}, {Enoto}, {Ho},
  {Gendreau}, \& {Arzoumanian}}]{kuiper20}
{Kuiper}, L., {Harding}, A.~K., {Enoto}, T., {et~al.} 2020, The Astronomer's
  Telegram, 13985, 1

\bibitem[{{Kuiper} \& {Hermsen}(2009{\natexlab{a}})}]{kuiper09a}
{Kuiper}, L., \& {Hermsen}, W. 2009{\natexlab{a}}, \aap, 501, 1031,
  \dodoi{10.1051/0004-6361/200811580}

\bibitem[{{Kuiper} \& {Hermsen}(2009{\natexlab{b}})}]{kuiper09}
---. 2009{\natexlab{b}}, The Astronomer's Telegram, 2151

\bibitem[{{Kuiper} {et~al.}(2018){Kuiper}, {Hermsen}, \& {Dekker}}]{kuiper18}
{Kuiper}, L., {Hermsen}, W., \& {Dekker}, A. 2018, \mnras, 475, 1238,
  \dodoi{10.1093/mnras/stx3128}

\bibitem[{{Kumar} \& {Safi-Harb}(2008)}]{kumar08}
{Kumar}, H.~S., \& {Safi-Harb}, S. 2008, \apjl, 678, L43,
  \dodoi{10.1086/588284}

\bibitem[{{Kumar} {et~al.}(2021){Kumar}, {Shannon}, {Flynn}, {Os{\l}owski},
  {Bhandari}, {Day}, {Deller}, {Farah}, {Kaczmarek}, {Kerr}, {Phillips},
  {Price}, {Qiu}, \& {Thyagarajan}}]{kumarsub}
{Kumar}, P., {Shannon}, R.~M., {Flynn}, C., {et~al.} 2021, \mnras, 500, 2525,
  \dodoi{10.1093/mnras/staa3436}

\bibitem[{{Laha} {et~al.}(2020){Laha}, {Barthelmy}, {Cummings}, {Krimm},
  {Lien}, {Markwardt}, {Palmer}, {Sakamoto}, {Stamatikos}, \&
  {Ukwatta}}]{laha20}
{Laha}, S., {Barthelmy}, S.~D., {Cummings}, J.~R., {et~al.} 2020, GRB
  Coordinates Network, 28193, 1

\bibitem[{{Leahy} \& {Tian}(2008)}]{leahy08}
{Leahy}, D.~A., \& {Tian}, W.~W. 2008, \aap, 480, L25,
  \dodoi{10.1051/0004-6361:20079149}

\bibitem[{{Livingstone} {et~al.}(2006){Livingstone}, {Kaspi}, {Gotthelf}, \&
  {Kuiper}}]{livingstone06}
{Livingstone}, M.~A., {Kaspi}, V.~M., {Gotthelf}, E.~V., \& {Kuiper}, L. 2006,
  \apj, 647, 1286, \dodoi{10.1086/505570}

\bibitem[{{Livingstone} {et~al.}(2011){Livingstone}, {Ng}, {Kaspi}, {Gavriil},
  \& {Gotthelf}}]{livingstone11a}
{Livingstone}, M.~A., {Ng}, C.~Y., {Kaspi}, V.~M., {Gavriil}, F.~P., \&
  {Gotthelf}, E.~V. 2011, \apj, 730, 66, \dodoi{10.1088/0004-637X/730/2/66}

\bibitem[{Lorimer \& Kramer(2005)}]{handbookpulsar}
Lorimer, D., \& Kramer, M. 2005, Handbook of Pulsar Astronomy, Cambridge
  Observing Handbooks for Research Astronomers (Cambridge University Press)

\bibitem[{{Luo} {et~al.}(2021){Luo}, {Ransom}, {Demorest}, {Ray}, {Archibald},
  {Kerr}, {Jennings}, {Bachetti}, {van Haasteren}, {Champagne}, {Colen},
  {Phillips}, {Zimmerman}, {Stovall}, {Lam}, \& {Jenet}}]{luo21}
{Luo}, J., {Ransom}, S., {Demorest}, P., {et~al.} 2021, \apj, 911, 45,
  \dodoi{10.3847/1538-4357/abe62f}

\bibitem[{{Majid} {et~al.}(2020){Majid}, {Pearlman}, {Prince}, {Enoto},
  {Arzoumanian}, {Gendreau}, {Naudet}, {Kocz}, {Horiuchi}, {Harding}, {Ho}, \&
  {Kuiper}}]{majid20}
{Majid}, W.~A., {Pearlman}, A.~B., {Prince}, T.~A., {et~al.} 2020, The
  Astronomer's Telegram, 13988, 1

\bibitem[{{McLaughlin} {et~al.}(2009){McLaughlin}, {Lyne}, {Keane}, {Kramer},
  {Miller}, {Lorimer}, {Manchester}, {Camilo}, \& {Stairs}}]{rratredetected}
{McLaughlin}, M.~A., {Lyne}, A.~G., {Keane}, E.~F., {et~al.} 2009, \mnras, 400,
  1431, \dodoi{10.1111/j.1365-2966.2009.15584.x}

\bibitem[{{Mickaliger} {et~al.}(2020){Mickaliger}, {Rajwade}, {Stappers},
  {Lyne}, \& {Preston}}]{mickaliger2020}
{Mickaliger}, M.~B., {Rajwade}, K., {Stappers}, B., {Lyne}, A., \& {Preston},
  L.~L. 2020, The Astronomer's Telegram, 13950, 1

\bibitem[{{NASA High Energy Astrophysics Science Archive Research Center
  (HEASARC)}(2014)}]{heasoft14}
{NASA High Energy Astrophysics Science Archive Research Center (HEASARC)}.
  2014, {HEAsoft: Unified Release of FTOOLS and XANADU}.
\newblock \doeprint{1408.004}

\bibitem[{{Nice} {et~al.}(2015){Nice}, {Demorest}, {Stairs}, {Manchester},
  {Taylor}, {Peters}, {Weisberg}, {Irwin}, {Wex}, \& {Huang}}]{nds+15}
{Nice}, D., {Demorest}, P., {Stairs}, I., {et~al.} 2015, {Tempo: Pulsar timing
  data analysis}, Astrophysics Source Code Library, record ascl:1509.002.
\newblock \doeprint{1509.002}

\bibitem[{{Pedregosa} {et~al.}(2011){Pedregosa}, {Varoquaux}, {Gramfort},
  {Michel}, {Thirion}, {Grisel}, {Blondel}, {M{\"u}ller}, {Nothman}, {Louppe},
  {Prettenhofer}, {Weiss}, {Dubourg}, {Vanderplas}, {Passos}, {Cournapeau},
  {Brucher}, {Perrot}, \& {Duchesnay}}]{pedregosa11}
{Pedregosa}, F., {Varoquaux}, G., {Gramfort}, A., {et~al.} 2011, Journal of
  Machine Learning Research, 12, 2825, \dodoi{10.48550/arXiv.1201.0490}

\bibitem[{{Rajwade} {et~al.}(2022){Rajwade}, {Stappers}, {Lyne}, {Shaw},
  {Mickaliger}, {Liu}, {Kramer}, {Desvignes}, {Karuppusamy}, {Enoto},
  {G{\"u}ver}, {Hu}, \& {Surnis}}]{rajwade22}
{Rajwade}, K.~M., {Stappers}, B.~W., {Lyne}, A.~G., {et~al.} 2022, \mnras, 512,
  1687, \dodoi{10.1093/mnras/stac446}

\bibitem[{{Rasmussen} \& {Williams}(2006)}]{rasmussen06a}
{Rasmussen}, C.~E., \& {Williams}, C. K.~I. 2006, {Gaussian Processes for
  Machine Learning}

\bibitem[{{Rea} {et~al.}(2010){Rea}, {Esposito}, {Turolla}, {Israel}, {Zane},
  {Stella}, {Mereghetti}, {Tiengo}, {G{\"o}tz}, {G{\"o}{\u g}{\"u}{\c s}}, \&
  {Kouveliotou}}]{rea10}
{Rea}, N., {Esposito}, P., {Turolla}, R., {et~al.} 2010, Science, 330, 944

\bibitem[{{Straal} {et~al.}(2023){Straal}, {Gelfand}, \& {Eagle}}]{straal23}
{Straal}, S.~M., {Gelfand}, J.~D., \& {Eagle}, J.~L. 2023, \apj, 942, 103,
  \dodoi{10.3847/1538-4357/aca1a9}

\bibitem[{{Terlouw} \& {Vogelaar}(2014)}]{KapteynPackage}
{Terlouw}, J.~P., \& {Vogelaar}, M.~G.~R. 2014, {Kapteyn Package, version 3.4},
  {Kapteyn Astronomical Institute}, Groningen

\bibitem[{{Uzuner} {et~al.}(2023){Uzuner}, {Keskin}, {Kaneko},
  {G{\"o}{\u{g}}{\"u}{\c{s}}}, {Roberts}, {Lin}, {Baring}, {G{\"u}ng{\"o}r},
  {Kouveliotou}, {van der Horst}, \& {Younes}}]{uzuner2023}
{Uzuner}, M., {Keskin}, {\"O}., {Kaneko}, Y., {et~al.} 2023, \apj, 942, 8,
  \dodoi{10.3847/1538-4357/aca482}

\bibitem[{{Verbiest} {et~al.}(2012{\natexlab{a}}){Verbiest}, {Weisberg},
  {Chael}, {Lee}, \& {Lorimer}}]{verbiest2012}
{Verbiest}, J.~P.~W., {Weisberg}, J.~M., {Chael}, A.~A., {Lee}, K.~J., \&
  {Lorimer}, D.~R. 2012{\natexlab{a}}, \apj, 755, 39,
  \dodoi{10.1088/0004-637X/755/1/39}

\bibitem[{{Verbiest} {et~al.}(2012{\natexlab{b}}){Verbiest}, {Weisberg},
  {Chael}, {Lee}, \& {Lorimer}}]{verbiest12}
---. 2012{\natexlab{b}}, \apj, 755, 39, \dodoi{10.1088/0004-637X/755/1/39}

\bibitem[{Virtanen {et~al.}(2020)Virtanen, Gommers, Oliphant, Haberland, Reddy,
  Cournapeau, Burovski, Peterson, Weckesser, Bright, {van der Walt}, Brett,
  Wilson, Millman, Mayorov, Nelson, Jones, Kern, Larson, Carey, Polat, Feng,
  Moore, {VanderPlas}, Laxalde, Perktold, Cimrman, Henriksen, Quintero, Harris,
  Archibald, Ribeiro, Pedregosa, {van Mulbregt}, \& {SciPy 1.0
  Contributors}}]{scipy2020}
Virtanen, P., Gommers, R., Oliphant, T.~E., {et~al.} 2020, Nature Methods, 17,
  261, \dodoi{10.1038/s41592-019-0686-2}

\bibitem[{{Wang} {et~al.}(2022){Wang}, {Zhang}, {Dai}, \& {Cheng}}]{wang2022}
{Wang}, F.~Y., {Zhang}, G.~Q., {Dai}, Z.~G., \& {Cheng}, K.~S. 2022, Nature
  Communications, 13, 4382, \dodoi{10.1038/s41467-022-31923-y}

\bibitem[{{Wilms} {et~al.}(2000){Wilms}, {Allen}, \& {McCray}}]{wilms00}
{Wilms}, J., {Allen}, A., \& {McCray}, R. 2000, \apj, 542, 914

\end{thebibliography}
%%%%%%%%%%%%%%% TABLE %%%%%%%%%%%%%%%%%%%%%%%%%%%%%
\startlongtable
\begin{deluxetable*}{ccccc}
\centering
\tablecaption{Observation log\label{Tab:Table6}}
\tablecolumns{5}
\tablewidth{0pt}
\tablehead{
\colhead{Telescope} &
\colhead{Obs ID} &
\colhead{Start Day} &
\colhead{Exposure\tablenotemark{\footnotesize a}} &
\colhead{Pulsed flux\tablenotemark{\footnotesize b}}\\
\colhead{} & 
\colhead{} & 
\colhead{YYYY-MM-DD} &
\colhead{(ks)} & 
\colhead{($10^{-11}$ erg cm$^{-2}$ s$^{-1}$)} \\
}
\startdata
\nicer & 1033290117  & 2018-08-22   & 1.4 & 0.9 $\pm$ 0.6\\
\nicer & 1033290133  & 2018-11-10   & 5.9 & 0.4 $\pm$ 0.2\\
\nicer & 1033290135  & 2018-11-14   & 1.2 & 0.9 $\pm$ 0.4\\
\nicer & 1033290139  & 2018-11-18   & 1.3 & 0.6 $\pm$ 0.4\\
%%%%%%%%%%%%%%%%%%%%%%%%%%%%%%%%%%%%%%%%%%%%%%%%%%%%%%%%%%%%%%%%
\nicer & 2516010101  & 2019-03-10   & 6.9 & 0.3 $\pm$ 0.2\\
\nicer & 2516010301  & 2019-03-13   & 4.9 & 0.3 $\pm$ 0.2\\
\nicer & 2516010401  & 2019-03-17   & 3.1 & 0.6 $\pm$ 0.3\\
\nicer & 2516010601  & 2019-05-06   & 4.4 & 0.5 $\pm$ 0.2\\
\nicer & 2516010801  & 2019-07-06   & 1.1 & $<2$\\
\nicer & 2516011102  & 2019-09-24   & 2.9 & 0.8 $\pm$ 0.5\\
\nicer & 2516011301  & 2019-09-28   & 2.6 & 0.9 $\pm$ 0.4\\
\nicer & 2516011701  & 2019-11-13   & 2.3 & 0.7 $\pm$ 0.4\\
\nicer & 2516011407  & 2019-02-14   & 2.8 & 0.9 $\pm$ 0.4\\
%%%%%%%%%%%%%%%%%%%%%%%%%%%%%%%%%%%%%%%%%%%%%%%%%%%%%%%%%%%%%%%%
\nicer & 2516011406  & 2020-02-13   & 4.3 & 0.7 $\pm$ 0.3\\
\nicer & 2516011408  & 2020-02-15   & 6.3 & 0.4 $\pm$ 0.2\\
\nicer & 2516011502  & 2020-02-27   & 2.2 & $<1$\\
\nicer & 2516011503  & 2020-02-27   & 5.7 & 0.5 $\pm$ 0.3\\
\nicer & 3598010101  & 2020-03-26   & 6.1 & 0.5 $\pm$ 0.2\\
\nicer & 3598010302  & 2020-03-29   & 6.8 & 0.6 $\pm$ 0.2\\
\nicer & 3598010401  & 2020-04-02   & 2.6 & 0.6 $\pm$ 0.3\\
\nicer & 3598010402  & 2020-04-03   & 1.8 & 0.6 $\pm$ 0.4\\
\nicer & 3598010501  & 2020-04-29   & 9.4 & 0.6 $\pm$ 0.2\\
\nicer & 3598010601  & 2020-05-27   & 9.6 & 0.4 $\pm$ 0.2\\
\nicer & 3598010701  & 2020-06-26   & 7.3 & 1.6 $\pm$ 0.2\\
%%%%%%%%%%%%%%%%%%%%%%%%%%%%%%%%%%%%%%%%%%%%%%%%%%%%%%%%%%%%%%%%
\nicer & 3598010801  & 2020-07-25   & 3.8 & 3.7 $\pm$ 0.3\\
\nicer & 3598010802  & 2020-07-25   & 4.3 & 3.6 $\pm$ 0.2\\
\nicer & 3033290101  & 2020-08-02   & 1.7 & 2.8 $\pm$ 0.4\\
\nicer & 3033290102  & 2020-08-02   & 9.0 & 2.5 $\pm$ 0.2\\
\swift & 00032031217 & 2020-08-02   & 4.0 & 2.9 $\pm$ 0.8\\
\nustar & 80602315002 & 2020-08-05  & 41.6 & 1.6 $\pm$ 0.2\\
\nicer & 3033290103  & 2020-08-05   & 5.2 & 2.5 $\pm$ 0.2\\
\swift & 00032031218 & 2020-08-05   & 2.5 & 2.0 $\pm$ 0.9 \\
\nicer & 3033290104  & 2020-08-06   & 5.4 & 2.2 $\pm$ 0.2\\
\nicer & 3033290105  & 2020-08-06   & 5.2 & 2.2 $\pm$ 0.2\\
\nicer & 3033290106  & 2020-08-08   & 4.7 & 2.3 $\pm$ 0.2\\
\nicer & 3033290107  & 2020-08-08   & 6.9 & 2.3 $\pm$ 0.2\\
\nicer & 3033290108  & 2020-08-10   & 6.0 & 2.1 $\pm$ 0.2\\
\nicer & 3033290109  & 2020-08-11   & 5.2 & 2.1 $\pm$ 0.2\\
\nicer & 3033290110  & 2020-08-12   & 4.5 & 2.2 $\pm$ 0.2\\
\swift & 00032031219 & 2020-08-12   & 3.0 & 1.9 $\pm$ 0.8\\
\nicer & 3033290111  & 2020-08-13   & 3.3 & 2.1 $\pm$ 0.2\\
\nicer & 3033290112  & 2020-08-14   & 2.2 & 1.8 $\pm$ 0.3\\
\nicer & 3033290113  & 2020-08-17   & 3.5 & 2.4 $\pm$ 0.2\\
\nicer & 3033290114  & 2020-08-18   & 2.0 & 2.3 $\pm$ 0.3\\
\nicer & 3033290115  & 2020-08-19   & 3.9 & 2.0 $\pm$ 0.2\\
\nustar & 80602315004 & 2020-08-20  & 56.2 & 2.5 $\pm$ 0.2 \\
\swift & 00032031221 & 2020-08-20   & 2.2 & 2.1 $\pm$ 1 \\ 
\nicer & 3033290116  & 2020-08-21   & 3.1 & 4.2 $\pm$ 0.3\\
\nicer & 3598010901  & 2020-08-24   & 8.7 & 2.3 $\pm$ 0.2\\
\nicer & 3033290117  & 2020-08-26   & 4.7 & 2.0 $\pm$ 0.2\\
\nicer & 3033290118  & 2020-08-28   & 3.3 & 1.9 $\pm$ 0.3\\
\nicer & 3033290119  & 2020-08-30   & 3.8 & 1.5 $\pm$ 0.3 \\
\nicer & 3033290120  & 2020-09-01   & 3.1 & 1.7 $\pm$ 0.3 \\
\nicer & 3033290121  & 2020-09-03   & 1.3 & 1.4 $\pm$ 0.4 \\
\nicer & 3033290122  & 2020-09-05   & 4.7 & 1.5 $\pm$ 0.2 \\
\nicer & 3033290123  & 2020-09-07   & 4.5 & 2.2 $\pm$ 0.2 \\
\nicer & 3033290124  & 2020-09-09   & 2.2 & 1.6 $\pm$ 0.3 \\
\nicer & 3033290125  & 2020-09-11   & 2.2 & 1.5 $\pm$ 0.4 \\
\swift & 00032031225 & 2020-09-13   & 3.7 & 2.0 $\pm$ 0.8 \\
\nicer & 3033290126  & 2020-09-15   & 1.9 & 1.5 $\pm$ 0.4\\
\nustar & 80602315006 & 2020-09-17 & 78.1 & 0.8 $\pm$ 0.1\\
\xmm    & 0872990101 & 2020-09-17 & 9.9 & 1.4 $\pm$ 0.2\\
\nicer & 3033290127  & 2020-09-19   & 2.3 & 1.1 $\pm$ 0.3 \\
\nicer & 3033290128  & 2020-09-22   & 2.4 & 1.4 $\pm$ 0.3\\
\nicer & 3033290130  & 2020-09-30   & 2.4 & 1.0 $\pm$ 0.4\\
\xmm & 0872990301 & 2020-10-02 & 19.1 & 1.1 $\pm$ 0.1\\
\nicer & 3033290131  & 2020-10-03   & 2.9 & 1.2 $\pm$ 0.3 \\
\nicer & 3033290132  & 2020-10-05   & 1.8 & 1.2 $\pm$ 0.4 \\
\nicer & 3033290133  & 2020-10-07   & 1.0 & 2.0 $\pm$ 0.6 \\
\nicer & 3033290134  & 2020-10-07   & 0.6 & 1.8 $\pm$ 0.6 \\
\nustar & 80602315008 & 2020-10-09 & 153.4 & 0.7 $\pm$ 0.1\\
\nicer & 3033290135 & 2020-10-11 & 2.8 & 1.4 $\pm$ 0.3\\
\nicer & 3033290136 & 2020-10-13 & 1.8 & 1.3 $\pm$ 0.4\\
\nicer & 3033290137 & 2020-10-19 & 3.4 & 1.1 $\pm$ 0.3\\
\xmm & 0872990401 & 2020-10-23 & 33.9 & 1.1 $\pm$ 0.08\\	
\nicer & 3033290138 & 2020-10-27 & 1.5 & 1.0 $\pm$ 0.5\\
\nicer & 3033290139 & 2020-10-27 & 1.5 & 1.1 $\pm$ 0.5\\
\nicer & 3033290140 & 2020-10-29 & 2.6 & 1.0 $\pm$ 0.4\\
\nicer & 3033290141 & 2020-10-31 & 2.2 & 1.1 $\pm$ 0.4\\
\nicer & 3033290142 & 2020-11-02 & 1.7 & 0.7 $\pm$ 0.4\\
\nicer & 3033290143 & 2020-11-04 & 3.1 & 0.7 $\pm$ 0.3\\
%%%%%%%%%%%%%%%%%%%%%%%%%%%%%%%%%%%%%%%%%%%%%%%%%%%%%%%%%%%%%%%%
\nicer & 3033290144 & 2020-11-04 & 3.1 & 1.1 $\pm$ 0.4\\
\nicer & 3033290146 & 2020-11-09 & 3.0 & 0.9 $\pm$ 0.3\\
\nicer & 3033290147 & 2020-11-11 & 3.6 & $<0.5$\\
\nicer & 3033290148 & 2020-11-13 & 1.9 & 0.9 $\pm$ 0.4\\
\nicer & 3033290149 & 2020-11-14 & 1.9 & $<1$\\
\nicer & 3598011201 & 2020-11-18 & 1.0 & 0.9 $\pm 0.2$\\
\nicer & 3598011202 & 2020-11-21 & 3.1 & 0.7 $\pm 0.3$\\
%%%%%%%%%%%%%%%%%%%%%%%%%%%%%%%%%%%%%%%%%%%%%%%%%%%%%%%%%%%%%%%%
\nicer & 4607020101 & 2021-03-08 & 2.0 & 1.1 $\pm$ 0.3\\
\nicer & 4607020201 & 2021-03-09 & 2.4 & 0.9 $\pm$ 0.4\\
\nicer & 4607020203 & 2021-04-14 & 7.9 & 0.6 $\pm$ 0.2\\
\nicer & 4607020204 & 2021-04-15 & 4.2 & 0.9 $\pm$ 0.3\\
\nicer & 4607020301 & 2021-03-11 & 3.0 & 0.9 $\pm$ 0.3\\
\nicer & 4607020401 & 2021-04-17 & 2.0 & 0.6 $\pm$ 0.4\\
\nicer & 4607020501 & 2021-05-03 & 7.9 & 0.6 $\pm$ 0.2\\
\nicer & 4607020601 & 2021-06-08 & 7.5 & 0.6 $\pm$ 0.2\\
\nicer & 4607020701 & 2021-07-14 & 7.0 & 0.5 $\pm$ 0.2\\
\nicer & 4607020802 & 2021-07-23 & 6.6 & 0.5 $\pm$ 0.2\\
\nicer & 4607020804 & 2021-08-09 & 3.6 & 0.5 $\pm$ 0.3\\
%%%%%%%%%%%%%%%%%%%%%%%%%%%%%%%%%%%%%%%%%%%%%%%%%%%%%%%%%%%%%%%%
\enddata
\tablecomments{
\tablenotetext{a}{Including only the good time interval.} 
\tablenotetext{b}{Excluding PWN and SNR emission and in the energy range of 1-12 keV.} 
}
\label{Tab:Table5}
\end{deluxetable*}
%%%%%%%%%%%%%%%%%%%%%%%%%%%%%%%%%%%%%%%%%%%%%%%%%%%%%
\end{document}